\documentclass[acmtog,authorversion]{myarxiv}

\pdfoutput=1

\usepackage{natbib}
\usepackage{booktabs} 
\usepackage{adjustbox}
\usepackage[T1]{fontenc}
\usepackage[utf8]{inputenc}
\usepackage{microtype}
\usepackage{xspace}

\setcopyright{rightsretained}

\acmJournal{TOG}

\copyrightyear{2020}

\definecolor{maroon}{HTML}{500000}

\acmSubmissionID{341}

\begin{document}
\title[Digesting the Elephant]{Digesting the Elephant
  ---
  Experiences with
  Interactive Production Quality Path Tracing of the Moana Island Scene
} 


\author{
  Ingo Wald$^{1,2}$
  \quad
  Bruce Cherniak$^{1}$
  \quad
  Will Usher$^{3,1}$
  \quad
  Carson Brownlee$^{1}$
  \quad
  Attila \'Afra$^{1}$
  \quad
  Johannes G\"unther$^{1}$
  \quad
  Jefferson Amstutz$^{1}$
  \\
  Tim Rowley$^{1}$
  \quad
  Valerio Pascucci$^{3}$
  \quad
  Chris R. Johnson$^{3}$
  \quad
  Jim Jeffers$^{1}$
  \\
  $^{(1)}$Intel Corp\quad
  $^{(2)}$now at NVIDIA\quad
  $^{(3)}$SCI Institute, University of Utah
  }


\renewcommand{\shortauthors}{Wald et al.}

\begin{abstract}
  \textbf{Abstract}\\
New algorithmic and hardware developments over the past two decades have enabled
interactive ray tracing of small to modest sized scenes, and are
finding growing popularity in scientific visualization and games.
However, interactive ray tracing has not been as widely explored in
the context of production film rendering, where challenges
due to the complexity of the models and, from a practical standpoint,
their unavailability to the wider research community, have posed
significant challenges. The recent release of the Disney Moana Island
Scene has made one such model available to the community for experimentation.
In this paper, we detail the challenges posed by this scene to an
interactive ray tracer, and the solutions we have employed and developed
to enable interactive path tracing of the scene with full geometric and
shading detail, with the goal of providing insight and guidance
to other researchers.
\end{abstract}


\begin{teaserfigure}
  \centering
  \includegraphics[width=0.48\textwidth]{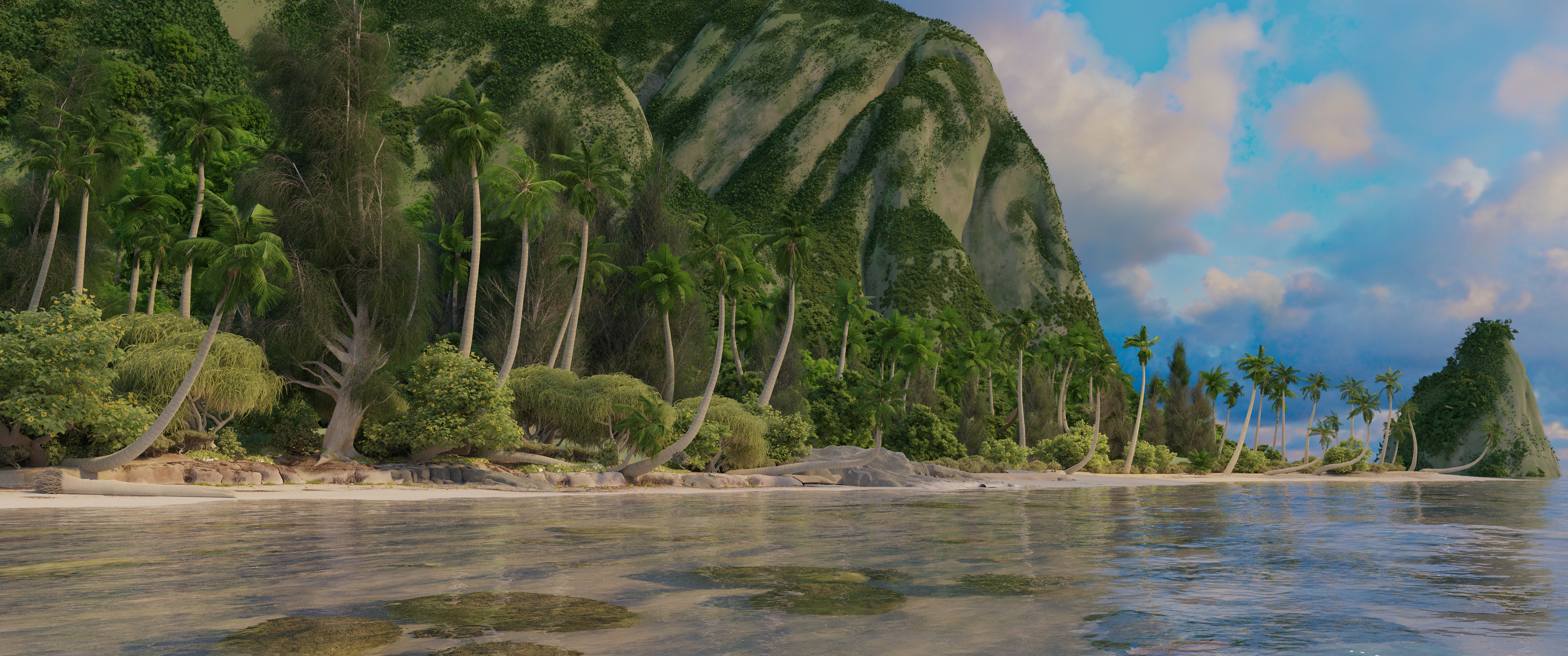}
  \includegraphics[width=0.48\textwidth]{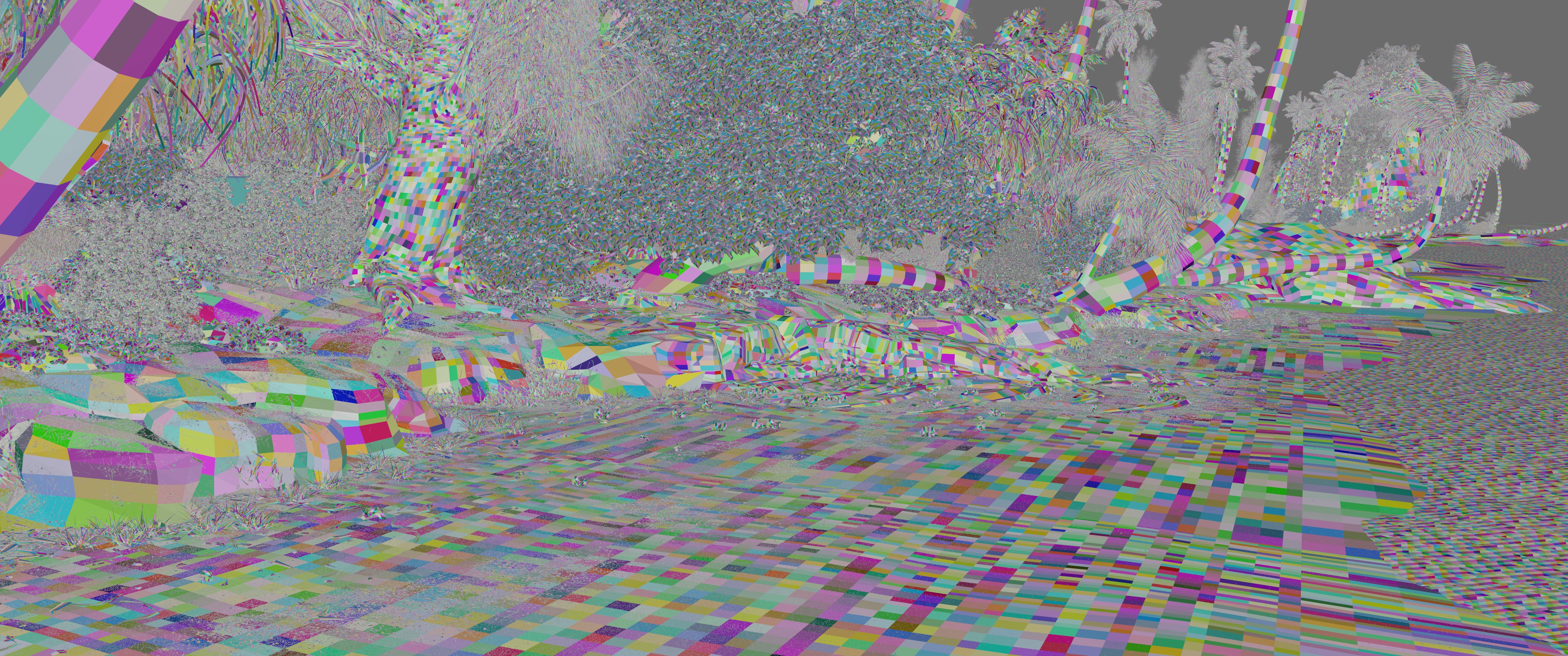}
  \\
  \vspace*{-1em}
	\caption{\label{fig:teaser}%
	Left: An overview of the Moana Island Scene (39.3
    million instances, 261.1 million unique quads, and 82.4 billion
    instanced quads). Right: A close-up of the beach, with
	each primitive colored individually to
    try to convey the detail of this model.
    In this paper, we describe the steps taken to allow us to render this
    model---in its entirety, without simplification, and with a high-quality
	path tracer---at interactive frame rates.}
\end{teaserfigure}
%
%


\keywords{Parallel Rendering, Path Tracing, Ray Tracing, Moana Island Scene}

\maketitle


\section{Introduction}
Over the past two decades rendering technology has seen an
astonishingly fast shift towards ray tracing.
While only a short time ago production movie
rendering predominantly used micro-polygon rasterization,
specifically Reyes~\cite{cook_reyes_1987},
today virtually all production renderers use some form of ray
tracing~\cite{pharr_tog_intro,keller_pt_course_2015,manuka,renderman,hyperion,arnold,arnold-sony}.
In the past few years even real-time rendering applications
have begun to incorporate ray tracing effects~\cite{battlefield-v,metro,picapica}.
This move toward ray tracing is driven by a number of factors:
ray tracing supports physically based lighting effects, which
improves artist usability and removes the need for approximations;
scales well with scene complexity; and maps well to thread parallelism.
With regard to the last two factors, significant improvements have
been made over the past two decades in both software and hardware to enable real-time ray tracing.


The push towards interactive ray tracing is most visible today in
scientific visualization applications~\cite{ospray,optix}
and games~\cite{battlefield-v,metro,picapica}, where
compelling ray traced images can be rendered at moderate
to real-time framerates on current CPUs and GPUs, depending
on the model size and shading complexity.
However, production movie rendering has largely remained an
offline process, taking minutes to hours per-frame.
Artist tools typically use rasterization for interactive rendering,
resulting in a disconnect between the lighting and material
models used in the modeling tools and the final renders.
This disparity in ray tracing performance is a result of the radically
different rendering demands of interactive applications and film.
Interactive applications are driven by hard real-time requirements, and
leverage a range of approximations or simplified models to produce convincing
imagery with just a few samples per-pixel;
however, production renderers are driven by
quality demands and prefer generality and true realism,
even if these require hundreds to thousands of samples per pixel.


The demands placed on the ray tracer further diverge
when considering the content each is tasked with rendering.
While game assets are modeled with a specific real-time frame
budget in mind, this is not the case for film, where assets are
re-used to save artist time or always modeled
at high-quality. For example, if a scene in a movie needs some
shells on a beach and a shell model is available, it is easiest
to re-use this asset, even if it was originally created for
a close-up shot and consists millions of triangles and high-resolution textures,
while the new use case may be for hundreds to thousands of sub-pixel objects.
Similarly, assets tend to be modeled at high-quality regardless
of their initial planned size on screen, as the shot may change to include
a close-up of the asset.


As a result, movie content is often detailed to a degree that
those outside of film may view as extreme, with many thousands
of triangles projecting to each pixel~\cite{manuka}, tens
to hundreds of gigabytes of textures, and orders of magnitude
more \emph{instances} than a typical game has \emph{triangles}.
To quote Matt Pharr's experience~\cite{matt_pharr_blog} in getting PBRT to
render Disney's Moana Island Scene,
dealing with such content is akin to ``swallowing an elephant'',
a surprisingly fitting description.


In this paper we follow Pharr's example, but with the explicit goal
of enabling \emph{interactive} rendering, without
compromises in model complexity or shading detail.
We describe our experience in both ``swallowing the elephant'',
i.e., loading and rendering it at all;
and ``digesting'' it, i.e., rendering it at interactive rates.  
We detail the challenges faced in terms of geometric
variety and complexity, model size, texturing and shading complexity,
and performance; and how we tackled these challenges to enable
interactive rendering.
Specifically, this paper aims to:
%
\begin{itemize}
	\item Detail the challenges posed by production film assets
		to an interactive ray tracer;
	\item Present our solutions for tackling these challenges
		to enable interactive rendering;
	\item Summarize what is possible today for interactive
		rendering of production assets, and briefly discuss
		future challenges.
\end{itemize}

%


\section{Paper Overview}
The goal of this paper is to detail the challenges encountered
and solutions developed in our efforts to enable interactive rendering of the
Moana Island Scene---exactly as provided by Disney, without simplification
and with production-style path traced image quality.

In Section~\ref{sec:related_work} we give a brief overview of related
work in production rendering and to the Moana Island Scene,
but defer specific related work
discussion to the relevant technical subsections throughout Section~\ref{sec:our-approach}.
To properly motivate the trade-offs and design decisions
made in our approach, it is imperative to
first convey the challenges posed by the Moana Island Scene. Though the
amount of detail in a modern movie shot is clear even to those
outside of graphics, few researchers outside of the studios
have had the chance to work with production content.
As such, the exact kind---and in particular scale---of the challenges
posed by production content is often not well understood in the broader
research community. To properly convey these challenges we
spend Section~\ref{sec:the-challenge} describing the scene, and the variety
and scale of challenges it poses to a ray tracer.
Though other production content may pose a somewhat different set
of challenges, the Moana Island Scene provides a good proxy for other
such content.

In Section~\ref{sec:our-approach} we then describe our approach
to handling these challenges, and discuss the individual
components of our final system, with a focus on those
which did not work out of the box, or had to be added specifically
for this work. In Section~\ref{sec:results} we present performance
results of our final system, and end with a brief discussion
and summary (Section~\ref{sec:discussion}).



\section{Related Work}
\label{sec:related_work}


Early versions of the Moana Island Scene were made available to researchers
as early as late 2017, and the first public version released in mid 2018.
An illuminating summary of some of the challenges in dealing
with the Moana Island Scene is available online through Matt
Pharr's series of blog posts~\cite{matt_pharr_blog} on ``Swallowing the Elephant'',
which discusses the challenges encountered in getting the model loaded and rendered
offline using PBRT~\cite{pbrt}.

In this paper we give a further in-depth discussion of the model
and the challenges it poses to a ray tracer, with a
focus on interactive ray tracing.
We build our interactive renderer on top of Embree~\cite{embree}
and OSPRay~\cite{ospray}, for ray traversal and rendering,
along with Disney Animation's \emph{Ptex} library~\cite{ptex}
and \emph{principled} BSDF~\cite{disney-principled,disney-principled-bsdf}, for texturing and shading.

Though the general inaccessibility of production assets outside
of the movie studios means that dealing with such assets is largely
uncharted territory for interactive rendering research, production renderers deal
with such content on a daily basis. An excellent survey of production
renderers was recently presented in the ACM Transactions on Graphics
Special Issue on Production Rendering~\cite{tog-special-issue},
which includes in-depth descriptions of some of today's most prominent
production renderers: Autodesk's Arnold~\cite{arnold}
and Sony's Arnold~\cite{arnold-sony},
Disney's Hyperion~\cite{hyperion}, Weta's Manuka~\cite{manuka}, and
Pixar's RenderMan~\cite{renderman}.
Many of these production renderers have added some support
for interactive model rendering (and in some cases, editing)
over the past few years, typically through
some form of progressive re-rendering of lower resolution
images when a part of the scene or viewpoint changes.




\section{The Challenge: The Moana Island Scene}
\label{sec:the-challenge}
The Moana Island Scene was publicly released in June 2018, and comes with an
extensive whitepaper describing the asset~\cite{moana-whitepaper},
which mentions that the publicly released version is only an
approximation of the original content (e.g., subdivision surfaces are represented
by their base cages).
The asset comes in two separate compressed archives,
together comprising a total of 51~GB of compressed data (134~GB uncompressed).
An additional archive specifying the animation data is also
provided, containg 24~GB of compressed data (131~GB uncompressed).

Even without animation data the model is 134~GBs, split
across roughly three components: a version of the model in a
proprietary JSON-encoding, a version of the model converted to PBRT
format, and textures which are shared between both versions.
The amount of data needed to render the scene is less than the total 134~GBs,
as only one of the JSON or PBRT versions are needed.

The scene contains 3749 Ptex textures, comprising a total of 41~GBs
of texture data. Of these textures
roughly 4\% are used for displacement mapping,
with the majority used for color mapping.

The JSON version of the scene contains 20~GB of JSON data across 165
files which specify the shading information and scene hierarchy, along
with an additional 11~GB of Wavefront OBJ files, which specify the polygon
meshes referenced by the JSON files. Virtually all polygons
are quads, and correspond to the subdivision base cages used in the
original scene.

The PBRT export of the scene contains 39~GB of data across 495~files.
There are a variety of differences between the PBRT and
JSON versions of the model, e.g., quad meshes are converted to
PBRT triangle meshes, materials are slightly different, etc., (for
more detail, see~\cite{moana-whitepaper}); however, the overall scene
structure remains the same.
For the remainder of this section we will
refer to the PBRT version of the model; the JSON version should
produce similar numbers, with the exception that every pair of triangles in
the PBRT files corresponds to a single quad in the JSON's OBJ files.

\subsection{Statistical Data}

In terms of shading data, there are 95 different PBRT
materials in the scene, all using Disney's principled
BSDF~\cite{disney-principled-bsdf}.
For many of these materials the
diffuse component is modulated by the underlying shape's
Ptex texture. Finally, there is one textured environment light
source and 23 quad-shaped key lights, which for a production asset is
a rather modest number.

\begin{figure}
  \centering
	\includegraphics[width=0.48\textwidth]{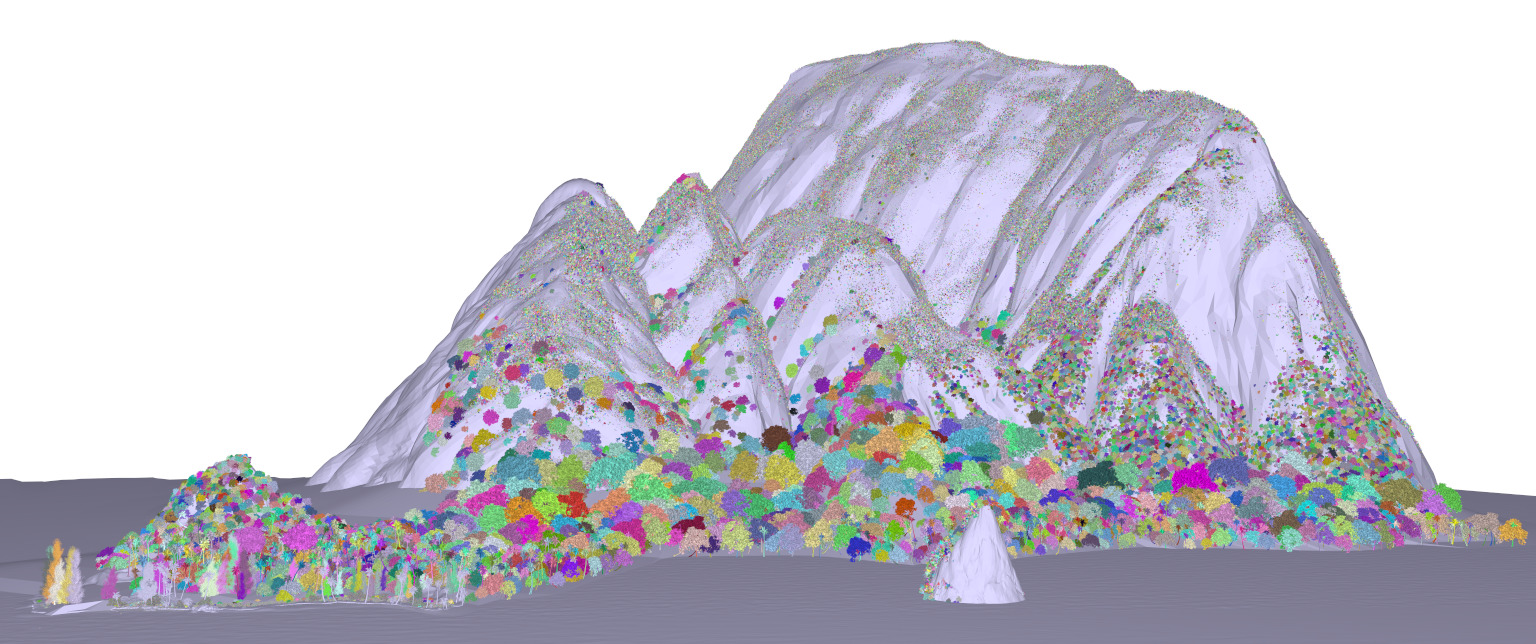}
  \\
  \vspace*{-1em}
  \caption{\label{fig:instances-overview}%
	An overview over the Moana
    Island Scene, with a pseudo-color ``instance ID'' shader to convey the
    large number of instances. Different object colors indicate different instances.
    \\
  \vspace*{-2em}
}
\end{figure}

As for geometry, there are a total of 278 unique PBRT ``objects''
(i.e., before instantiation), with a total of 1.9~M unique PBRT
``shapes'' (mostly triangle meshes) which contain a total of 146~M
unique triangles. After instantiation these correspond to 39~M
instanced objects with 106~M shapes, containing 10~M curves and 164
billion instanced triangles. There are an additional 375~K ``ribbon''
curves (some round, some flat) with roughly 10~M cubic curve segments.
These curves are primarily used for grass and some palm fronds in the scene.
To emphasize the scale of the geometry in the scene, the Moana Island contains
more geometry \emph{instances} (over 100~M) than most scenes used in ray tracing
research contain in final polygons (also see Figure~\ref{fig:instances-overview}).

\subsection{Beyond Statistics}

While the statistics behind the scene are already impressive,
raw statistical information does not convey the true challenges
posed by how this translates to the scene geometry.
An adequate description is best provided through visual exploration in an interactive
session (see supplemental video on YouTube~\footnote{\url{https://youtu.be/JHyC7DE3mJ4}}),
and in text form will necessarily fall short.
However, to at least convey some of these challenges: a frequently occurring
feature in the scene are significantly overlapping or intersecting geometric shapes, either
resulting from two physical copies of the same logical asset (in particular, trees),
or finer detail levels added on top of coarser base geometry.
The latter case is encountered on the water and island surface geometry,
where a finer top surface was placed on top of coarser base one.

\begin{figure}
  \centering
  \adjustbox{trim={{.2\width} 0 {.3\width} 0},clip}{
	  \includegraphics[width=0.98\columnwidth]{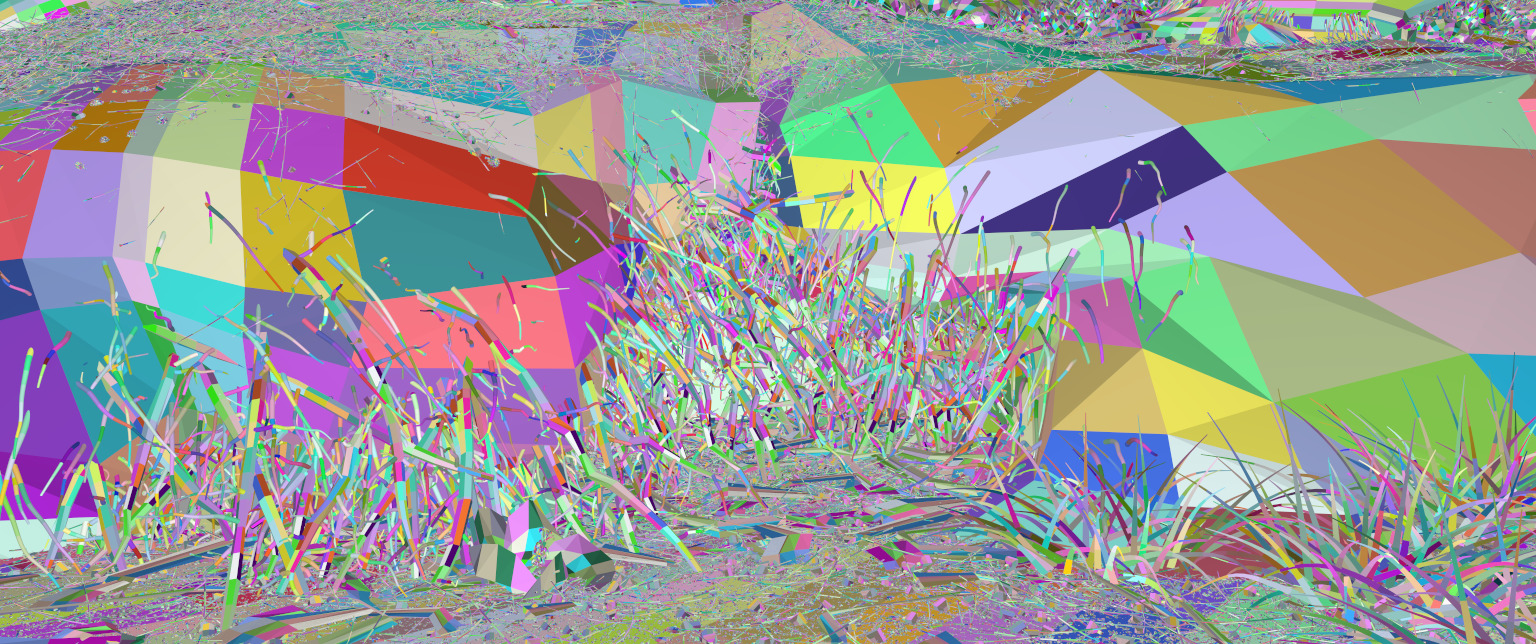}}
  \adjustbox{trim={{.2\width} 0 {.3\width} 0},clip}{
	  \includegraphics[width=0.98\columnwidth]{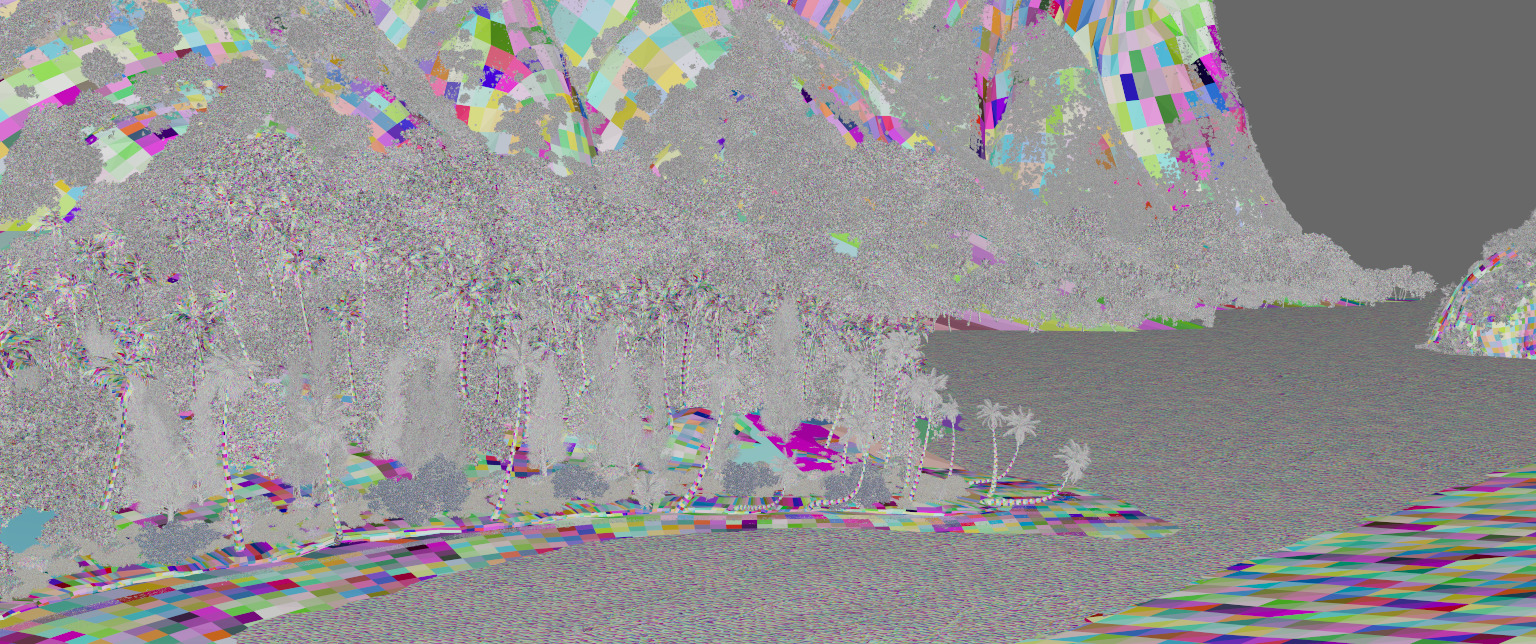}}
  \\
  \vspace*{-1em}
  \caption{\label{fig:varying-prim-density}%
	Near and far views of the Moana Island, with a pseudo-color ``primitive ID''
	shader to convey the highly varying geometric density. Left: Detailed
	individual twig, pebble and seed models sit on the coarser island terrain mesh.
	Right: The ocean surface near the land is highly tessellated, with the beach
	covered with the sub-pixel twigs, pebbles and seeds seen in the left.
  \\
  \vspace*{-2em}
  }
\end{figure}

Another common feature in the scene are large variations in the tessellation density
of nearby objects (Figure~\ref{fig:varying-prim-density}),
either due to the relative sizes of the objects, e.g., 1k+ triangle twigs
on a beach tessellated according to a mile-sized island, or
due to some view- and curvature-adaptive tessellation of the water.
The water surface alone is a few million triangles, with some areas
using almost millimeter-scale tessellation.
Some of these scale differences come from
instantiation (e.g., twigs, grains of sand), while some (e.g., the water) are in the base
geometry.
Large and tiny geometric primitives often overlap
and, combined with the large number of long, thin polygons (e.g., branches, roots),
this poses a significant challenge to the ray tracer's acceleration structure.


\begin{figure}
  \centering
  \adjustbox{trim={{.2\width} 0 {.3\width} 0},clip}{
	  \includegraphics[width=0.98\columnwidth]{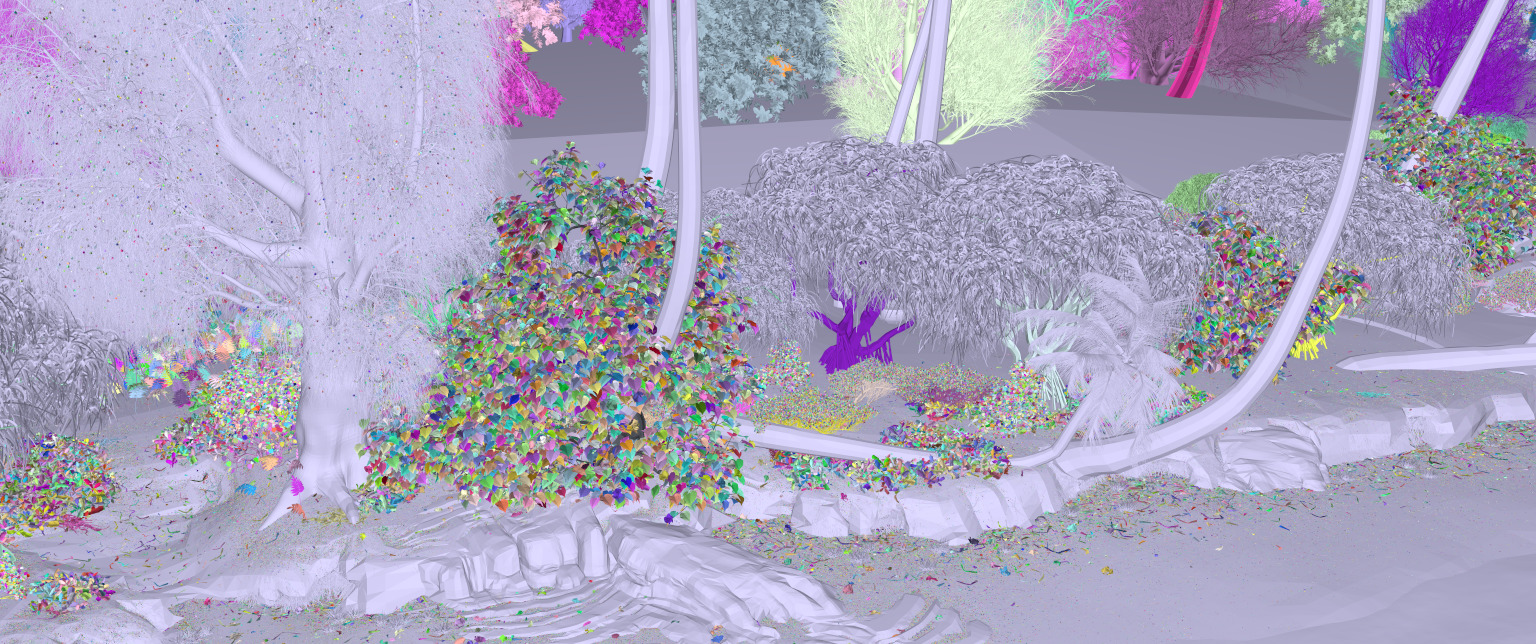}}
  \adjustbox{trim={{.2\width} 0 {.3\width} 0},clip}{
	  \includegraphics[width=0.98\columnwidth]{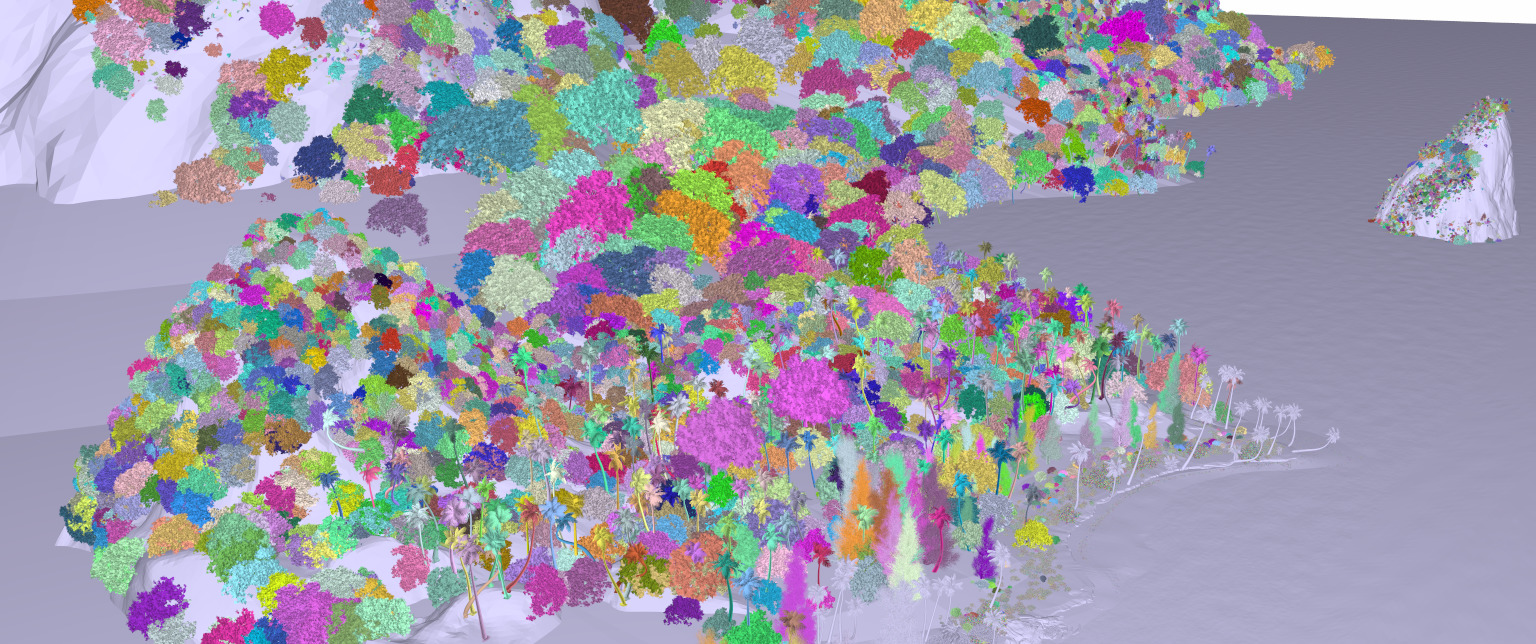}}
  \\
  \vspace*{-1em}
  \caption{\label{fig:instance-size-and-overlap}%
  The size of instances and large amount of overlap among them pose
  significant challenges to the ray tracer's top-level BVH.
  Left: Individual leaves of some bushes are instanced, and overlap significantly
  with each other and geometry from other instances.
  Right: The large number of trees instanced through the scene overlap
  significantly.
  \\
  \vspace*{-2em}
  }
\end{figure}

The distribution of instances in the scene poses a similarly challenging situation.
Some instances are large, both physically and in number of primitives,
such as some trees which consist of several million triangles. 
As the entire geometry for the tree is in a single BVH, these are simple
cases for the ray tracer. However, other instances are tiny,
consisting of a single leaf or flower bud made of a few dozen
triangles. These small instances are then used thousands of times, e.g.,
to place leaves on a bush, and significantly overlap each other and other scene geometry
(see~Figure~\ref{fig:instance-size-and-overlap}). Although each instance
is small, the extreme amount of overlap effectively disables the top-level
BVH's ability to separate these objects. Moreover, many instanced objects
in the scene (e.g., trees) contain a large amount of empty space, thus
there is a high chance they will have to be traversed by the ray
tracer, but a low chance of intersecting them, leading to
a significant increase in traversal cost (Figure~\ref{fig:instance-size-and-overlap}).

\begin{figure}
  \centering
  \adjustbox{trim={{.35\width} 0 {.15\width} 0},clip}{
	  \includegraphics[width=0.98\columnwidth]{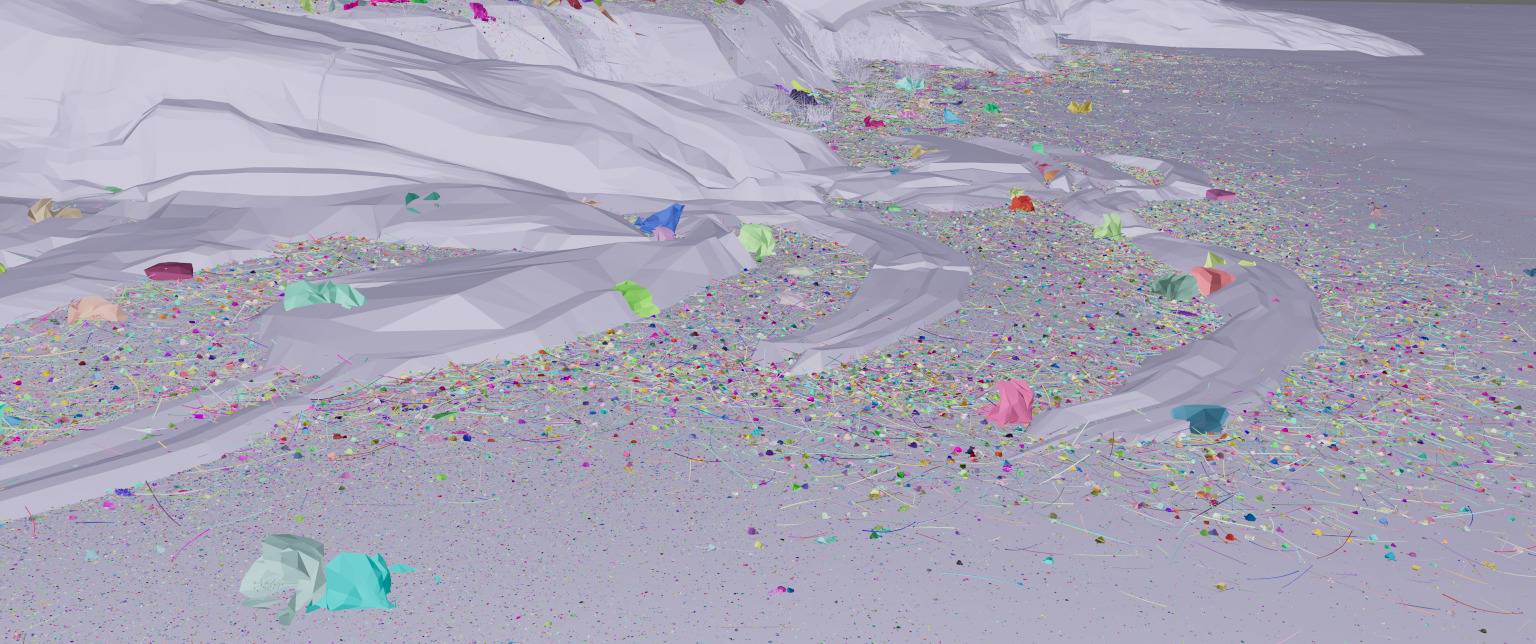}}
  \adjustbox{trim={{.2\width} 0 {.3\width} 0},clip}{
	  \includegraphics[width=0.98\columnwidth]{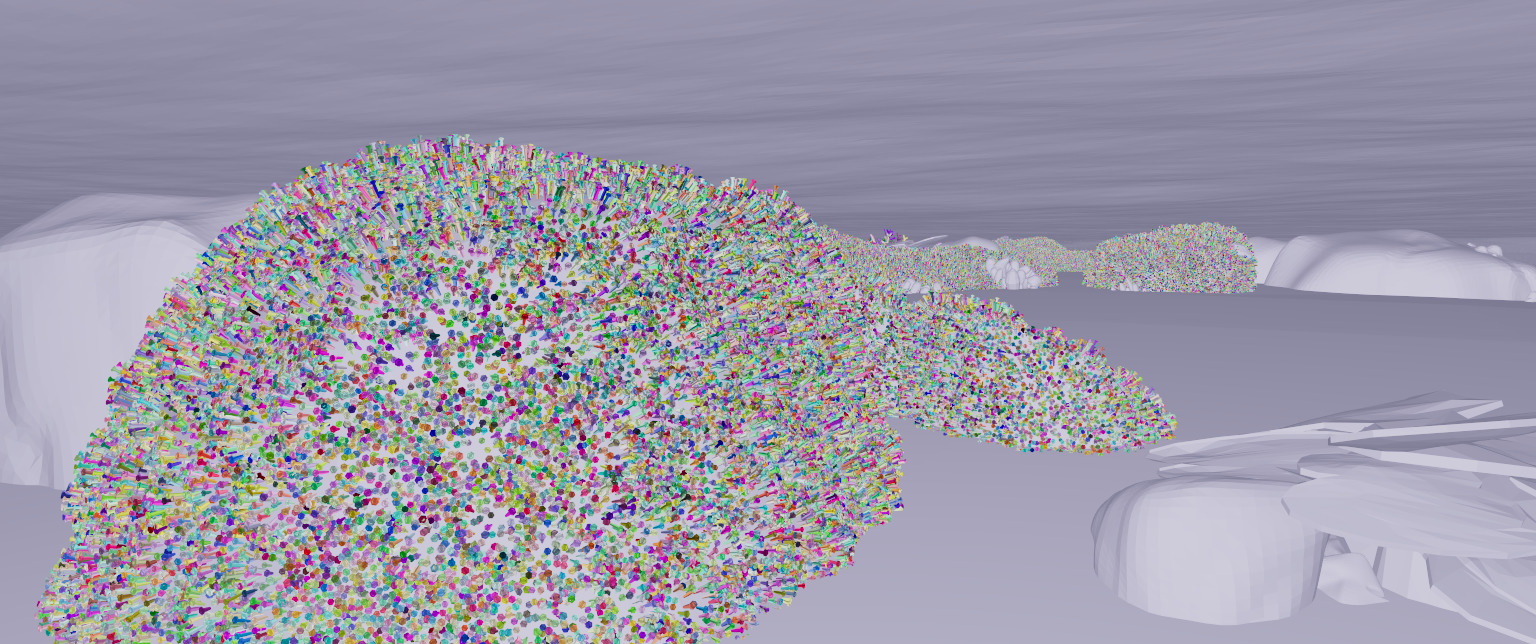}}
  \\
  \vspace*{-1em}
  \caption{\label{fig:pebbles-and-antlers}%
	Pebbles, twigs, and flowers (left); and coral antlers (right);
	are modeled at high detail and instanced, creating a massive
	amount of detailed sub-pixel and off-camera geometry.
  \\
  \vspace*{-1em}
  }
\end{figure}

Beyond the challenges the scene poses to a ray tracer,
the overall level of geometric detail in the model is difficult
to convey. For example, there are millions of object instances
modeled at high detail which for anything other than a close-up
shot will be sub-pixel, off-camera, or occluded in the final frame
(see~Figure~\ref{fig:pebbles-and-antlers}).
For those outside of production rendering it may be tempting
to dismiss such data as ``extreme'' or ``overmodeled''.
In practice, such assets are a natural consequence of a
workflow which prioritizes artist time, asset re-use, and realism,
where scenes are assembled from many smaller high-quality assets
and through content generation tools
such as Disney's Bonsai~\cite{disney-bonsai}.



\section{Digesting the Elephant}
\label{sec:our-approach}
From the beginning, our goal was to
enable interactive visual exploration of the Moana Island
Scene described in the previous section, at full geometric and shading quality.
Though we were initially confident that Embree~\cite{embree} and OSPRay~\cite{ospray}
could handle such content out of the box, we found this to not always be the case.
The challenges we encountered in this work can be roughly grouped into the
following five categories:
data wrangling (i.e., dealing with dozens of gigabytes of input data);
geometric complexity and variety (i.e., non-triangular primitives);
the use of Ptex textures throughout the model;
OSPRay's path tracer's inability to handle the model's shading
demands (i.e., the principled material);
and the need to achieve interactive performance.

\subsection{Data Wrangling}
An often overlooked but crucial component when working with large scenes
is the ability to load the data at all, in a reasonable time and memory budget. While not
an issue for smaller scenes, in the case of a production scene at the
scale of the Moana Island, just loading the data poses a real challenge.
An excellent discussion of some of these challenges can also be found in Matt Pharr's
``Swallowing the Elephant'' series of blog posts~\cite{matt_pharr_blog},
which covers the challenges of loading the Moana Island Scene into PBRT.

Given both the JSON and PBRT versions of the asset\footnote{We were
graciously provided some early versions before the public release, the latest
release of the asset also includes scripts to export an Embree XML version.}, we conducted
some early experiments using our open-source PBRT parser, though
quickly ran into its limitations. Attempts to load the original JSON
format directly proved similarly challenging. Though the whitepaper~\cite{moana-whitepaper}
provides some documentation about the JSON format, the nature of
JSON as a purely syntactical encoding makes it challenging to identify
semantic relationships between entities across the large number of
files and directories. This effort was further complicated by the fact
that there appears to be a form of multi-level instancing used
in the JSON version, though it was not always clear which directories
used instances in which way.
As just reading the 20~GBs of JSON
data could take minutes, with little to verify the parser's output,
debugging the JSON parser turned into a major issue.

%

Thus we returned to the PBRT version, and developed a working
PBRT parser using the public PBRT v3 scenes for verification.
Armed with this parser we returned to the Moana
Island Scene, and began testing on subsets of the scene by manually
editing the root \texttt{island.pbrt} to remove objects the parser
initially did not support (e.g., curves) as they were gradually added in.


While our PBRT parser was now able to load the scene,
parsing larger and larger subsets of the scene quickly led
to the parsing time becoming a huge bottleneck.
As PBRT is an ASCII format, a significant amount of time was spent
performing billions of \texttt{fscanf}s to read the geometry data.
To alleviate this issue, we developed an internal binary file format,
BIFF\footnote{A variant of the BIFF format
and our PBRT parser are available on GitHub, at~\url{https://github.com/ingowald/pbrt-parser/}},
which maintained the exact same structure
as the original PBRT files, but stored the geometry data in a binary
format which could be read directly with \texttt{fread} instead.
After converting the scene to BIFF, we were able to parse the
model (without textures) in just 15s, compared to 65 minutes using our PBRT
parser. The total scene load and setup time to create the geometries, build the BVHs
and so on takes approximately 6 minutes.

\subsection{Geometry Types, Complexity, and Memory Consumption}

With our parser now able to load the triangle meshes in the scene,
we loaded these into OSPRay and Embree to render them. Though the model
also contains 3~M B\'ezier curve segments, these are clearly
dwarfed by the 164 billion triangles (after instantiation),
and thus we began with just the triangle data.
Our initial tests rendered the meshes colored by primitive, geometry, or
instance ID using OSPRay's built in debug renderers
(see Figures~\ref{fig:instances-overview}-\ref{fig:pebbles-and-antlers}).


Though this approach worked reasonably well from a rendering standpoint,
it required a significant amount of memory just for the geometry.
Upon closer investigation, we found several ways
of reducing the memory consumed by the geometry.
First, as PBRT does not support quad primitives, the
JSON-to-PBRT converter exported each quad
as a pair of triangles, along with
filler texture coordinates and additional data arrays to remap the
triangle IDs to quad IDs for Ptex texturing. Removing these arrays
and computing these values on the fly provided a significant reduction in memory use.
Second, we initially instantiated PBRT \emph{geometries}
rather than \emph{objects}. While this was not a problem for
other PBRT models, on the Moana Island this approach significantly increased the
number of instances in the top-level BVH, from around 39~M to over 100~M.
Correcting the parser to instantiate PBRT objects instead provided
a further reduction in memory use.



\subsubsection*{Quads}
Having already partially reverted the quad to triangle-pair conversion by
computing the texture coordinates and quad IDs on the fly, the obvious next
step to reduce memory use further was to completely revert the
tessellation, and render quads directly.
From the content side this was straightforward, as a visual inspection
of the PBRT files revealed that every pair of triangles formed
a quad, and thus could simply be merged back together when converting
the data to BIFF.

On the rendering side, however, this proved more challenging.
While Embree had recently added a quad mesh primitive in version 3,
OSPRay was still on Embree 2, and did not support quads.
As OSPRay was initially developed for scientific visualization,
where quads are uncommon, this had not previously been an issue.
Adding support for rendering quads in OSPRay required upgrading
OSPRay to Embree 3, which, due to changes in how user geometry
work between Embree 2 and 3, and their extensive use throughout OSPRay,
was a significant effort.
After upgrading to Embree 3 we added a \texttt{QuadMesh} geometry
to OSPRay which directly mapped to Embree's \texttt{QuadMesh}.
Migrating to quads immediately halves the number of geometric
primitives, which also reduces the number of BVH nodes, and thus build
time and memory use. Upgrading to Embree 3 also
provided some additional upgrades to the underlying BVH as well,
improving performance further.


\subsubsection*{Curves}
As with the quads, we initially skipped loading the B\'ezier curves
in the scene as Embree 2, and thus OSPRay, did not support them.
Although OSPRay did have support for stream line geometries,
this geometry was designed for visualization applications and 
did not support smooth cubic curves nor flat ribbon style curves,
making it unsuitable for representing grass and palm fronds.
However, after upgrading to Embree 3 we were able to use the new
cubic curve types which were recently added to Embree independently from this work.
Similar to adding quad support, all that had to be done was add a new \texttt{Curves}
geometry to OSPRay, which used Embree's curve primitive internally.


As a result of our efforts to reduce memory consumption when
loading and rendering the scene the full model can be rendered
on a workstation with 128~GB of RAM.
As Embree and OSPRay both allow
for zero-copy sharing of the vertex, index, and other data arrays
with the application, passing the data to the renderer requires
no additional memory use; though Embree will require some additional
space to store the BVHs. After constructing the BVHs the
viewer requires a total of 100~GB to hold the geometry and acceleration structures,
and reaches a peak memory use of 104~GB during the BVH build.


\subsection{Ptex}
\label{sec:ptex}
Texture data is used heavily in
production rendering, and the Moana Island Scene is no exception:
the diffuse component of nearly every primitive in the scene comes
from a Ptex texture.
Although OSPRay already supported textures, it only supported
\emph{image} textures, however Ptex is a \emph{geometry} based
texture format baked on top of the underlying meshes~\cite{ptex}.
Not only does this mean there is no reasonable way these textures
could be converted to 2D images for use in OSPRay, but that OSPRay's
entire view of how textures can be applied to geometry---which was inherently
based on image textures---would have to change.

Previously, a texture in OSPRay would be given just the 2D UV coordinates
to be sampled and return back the computed color. However, in the case
of a Ptex texture, we also need the primitive ID (in Ptex terms, the ``face ID'')
to find the correct texture to sample, and the barycentric coordinates
of the primitive. Modifying OSPRay to pass
this data as well was straightforward: rather than
passing just the UV texture coordinates to the texture, we pass it the full
intersection information, which includes the primitive ID and barycentric coordinates.
However, all of OSPRay's rendering, shading, and texturing code is written
in ISPC, which operates on multiple shade points in SIMD. While
we now had the right data to pass to Ptex for each sample, the library does
not provide a SIMD interface which can be called directly from ISPC.

To call back into the Ptex library we wrote a C-callable shim function
which could be called from ISPC with the data for a single sample.
In ISPC we then serialize the SIMD execution over the active
vector lanes using ISPC's \texttt{foreach\_active} construct, and
call our shim with the corresponding sample for the lane.
While this does lose the advantage of ISPC's vectorization during
texture lookups, in a path tracer it is likely that different
vector lanes will sample different textures, impacting SIMD utilization
regardless.
Further investigation into the potential for an ISPC version of Ptex
which can take advantage of SIMD for texture lookups remains an interesting
direction for future work.

On the C++ side of the Ptex texture object we use a \texttt{PtexCache}
to load and cache the texture data, which is shared across all textures
in the scene. The cache helps reduce memory use
as the required texture data is loaded on demand,
though comes at the cost of poor performance when first starting the viewer
as frequently used textures are first loaded into the cache.

\subsection{Shading}
\begin{figure}
  \centering
	  \includegraphics[width=0.98\columnwidth]{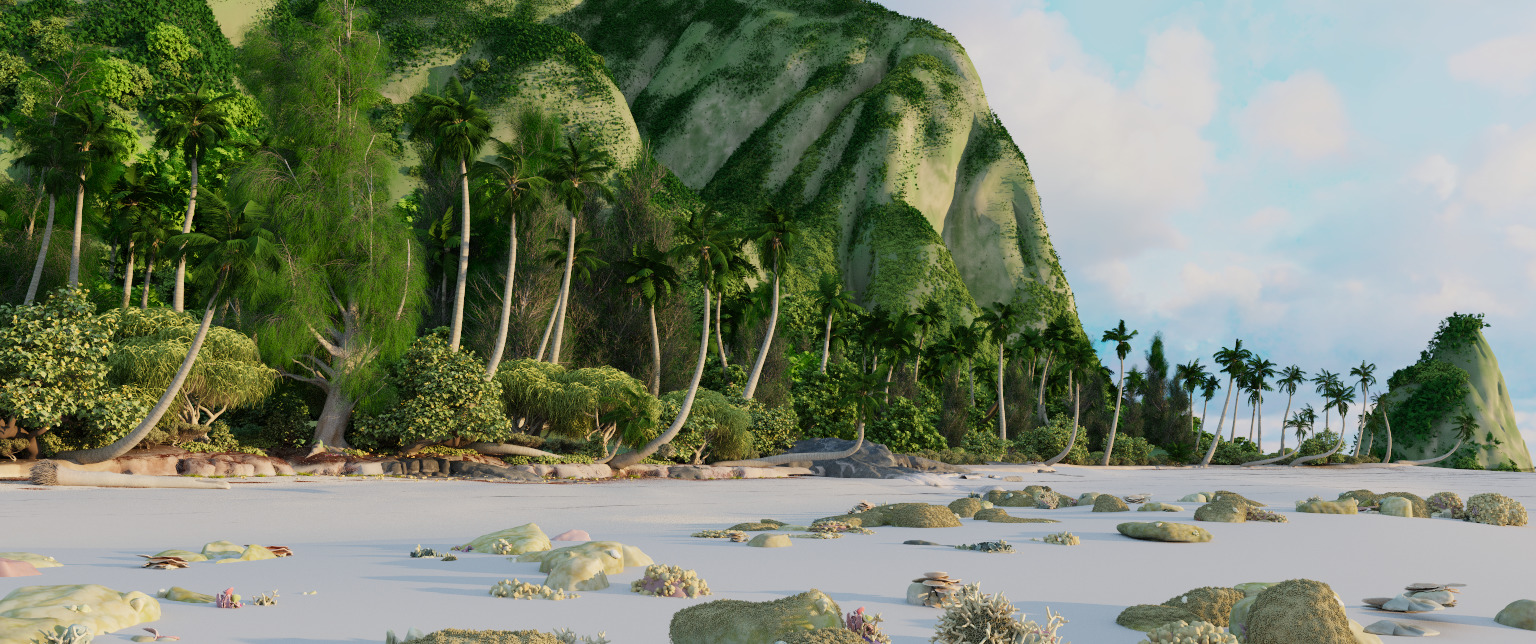}
	  \includegraphics[width=0.98\columnwidth]{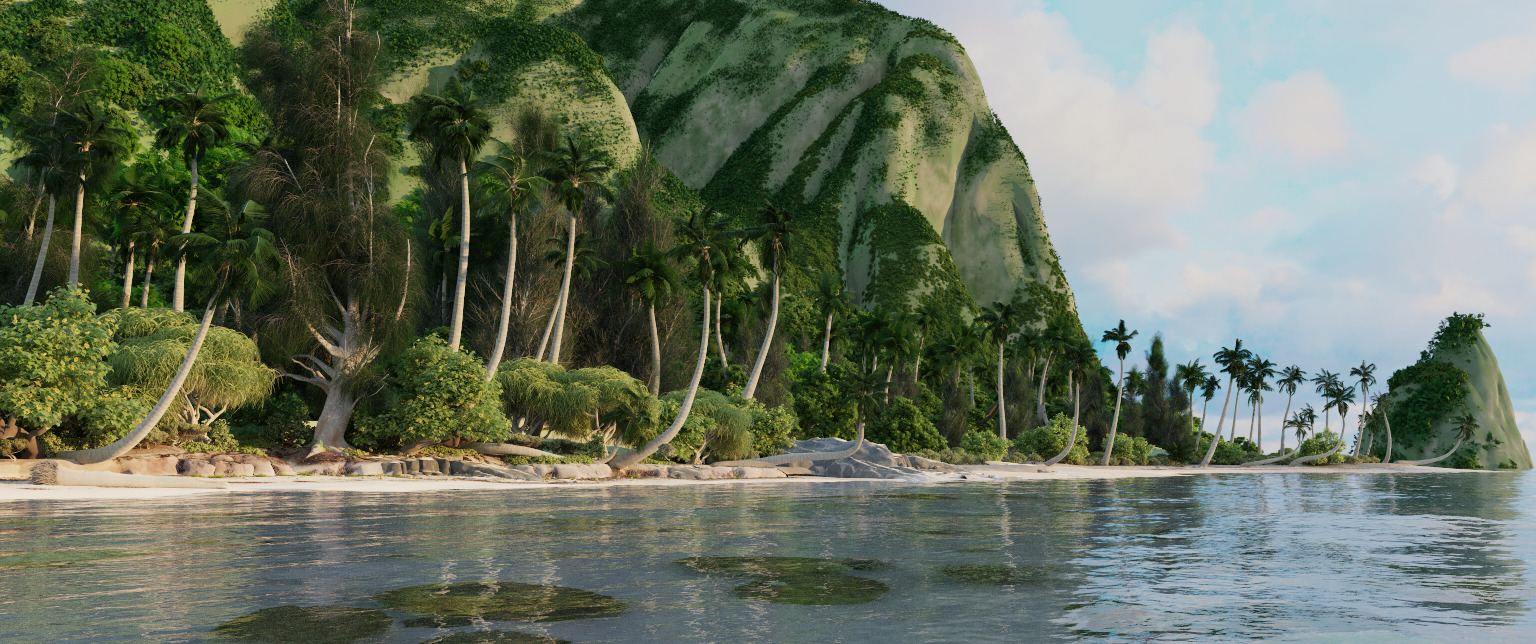}
  \\
  \vspace*{-1em}
  \caption{\label{fig:compare-obj-and-principled}%
	Top: Rendering with
    the simplified material model OSPRay used before this project.
    Bottom: The same, with the Disney principled material model that we
	added for this project. Note the incorrect colors on the Ironwood tree
	and the disappearance of the ocean in the top image.
  \\
  \vspace*{-1em}
  }
\end{figure}

Prior to this work, OSPRay had integrated
a reasonably full-featured path tracer. Though OSPRay was primarily
designed for scientific visualization, users' needs beyond
classical sci-vis had resulted in this path tracer evolving to support various
material types (e.g., glass, metal, plastic), different light
sources and area lights, performance optimizations for importance
sampling, and progressive refinement~\cite{ospray}.
However, our hope that this path tracer would meet the needs of the Moana
Island Scene out of the box was disappointed. The Moana Island Scene
exclusively uses the Disney principled BSDF, which could not be well
approximated by the existing materials in the path tracer.
Proper support for the principled BSDF was crucial to achieving
the correct look for the scene (see~Figure~\ref{fig:compare-obj-and-principled}),
and thus we have implemented a slightly modified and improved version of the
Disney BSDF in ISPC for OSPRay. One notable difference compared to the original
version of the BSDF is that we have made it both energy conserving \emph{and}
preserving~\cite{sony-pbs}.


\begin{figure}
  \centering
	  \includegraphics[width=0.98\columnwidth]{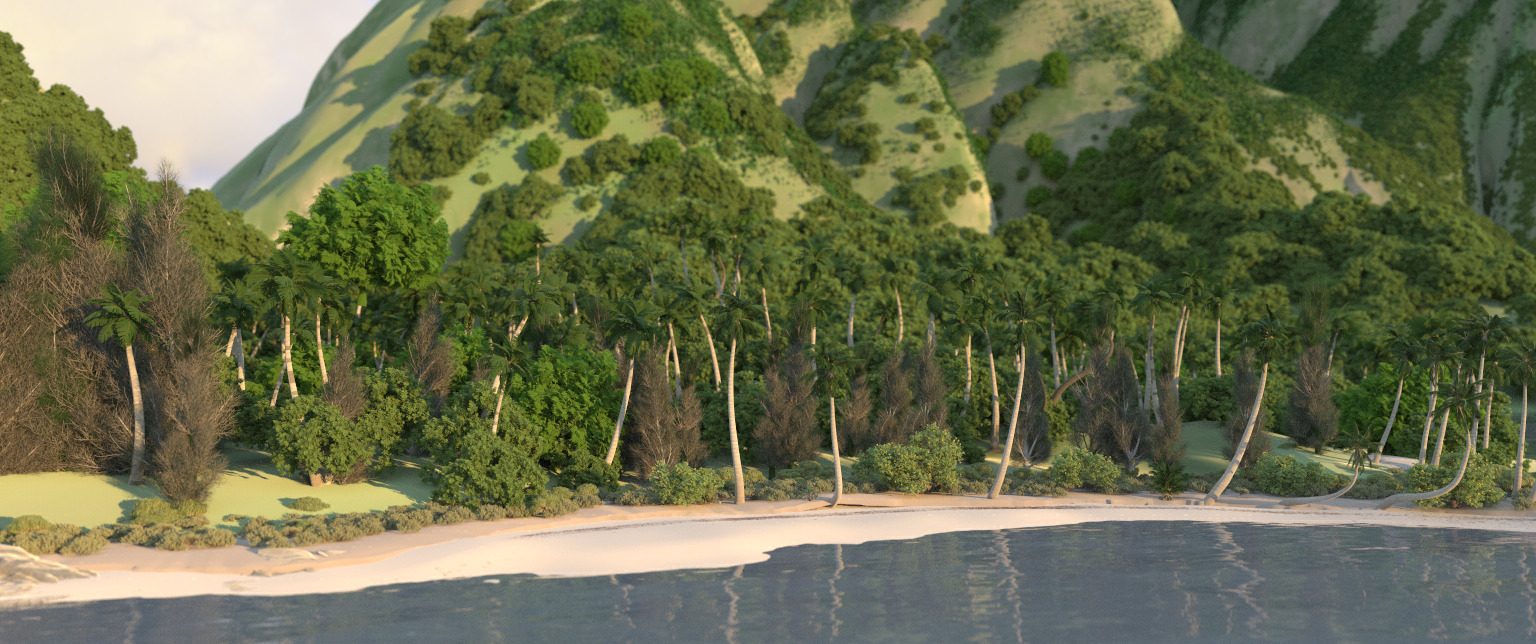}
	  \includegraphics[width=0.98\columnwidth]{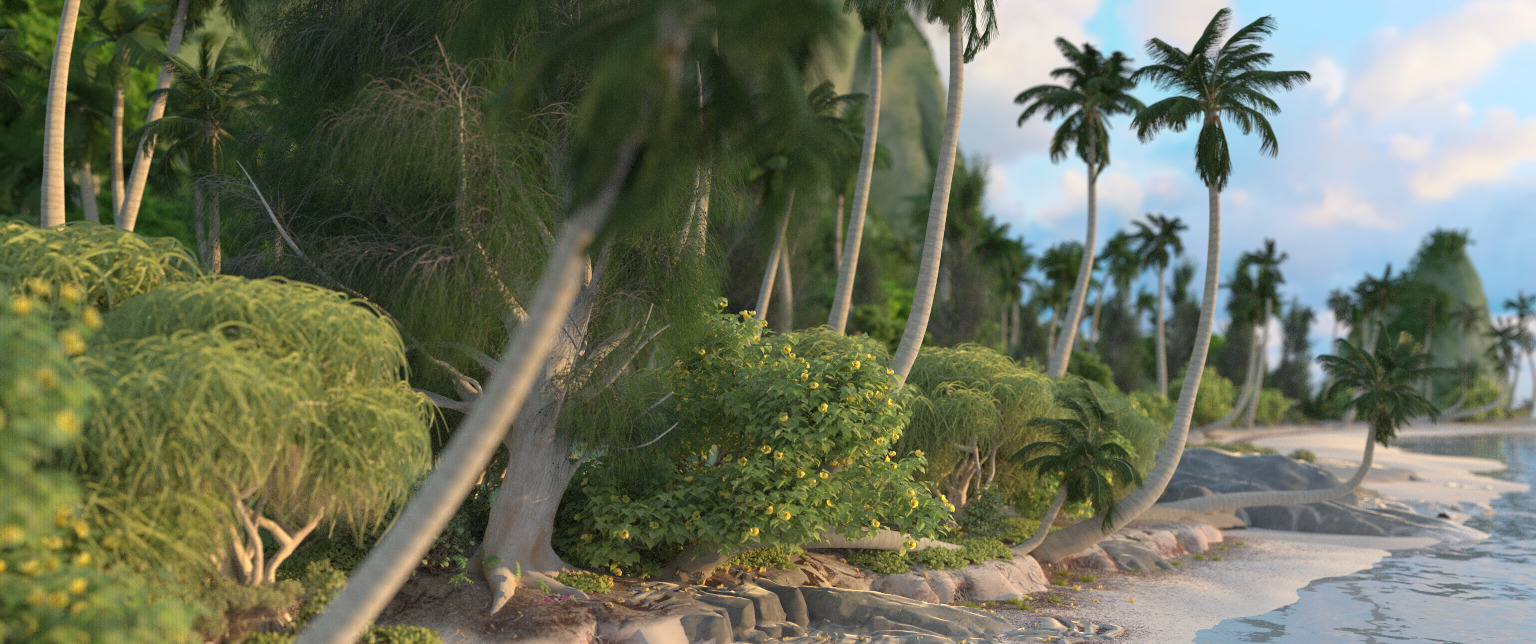}
  \\
  \vspace*{-1em}
  \caption{\label{fig:pt-final-beauty-shots}%
  The material model and path tracer support high-quality path tracing effects
	with progressive refinement, allowing interactive rendering of the Moana Island Scene
	with full geometric and shading quality. To provide interactive frame rates, we take one sample
	per-pixel each frame and accumulate these frames over time to refine the image.
    \\
  \vspace*{-1em}
}
\end{figure}

In its final version the path tracer can largely provide everything
the Moana Island Scene requires, achieving the desired look at reasonable
efficiency (see Figures~\ref{fig:teaser} and \ref{fig:pt-final-beauty-shots}).
As with any path tracer there is
potential for even better sampling, importance sampling, filtering,
etc. In particular, scenes with many more light sources than the Moana Island
would likely require additional support for improved sampling strategies.

\subsection{Performance}
\label{sec:performance}
OSPRay achieves interactive performance when rendering
the Moana Island Scene by leveraging a set of high-performance libraries,
code, and components. To leverage the SIMD capabilities on a single core OSPRay
uses Embree~\cite{embree}, which is vectorized internally, for ray intersections,
and implements the remainder of the renderer, material, and texture sampling code
in ISPC~\cite{ispc}.

On a single machine (or node in HPC terminology) OSPRay uses Intel's
Thread Building Blocks (TBB) for multi-threading, which provides utilities
for parallel for loops and asynchronous tasks. Work is distributed among
multiple threads by parallelizing the rendering task over the tiles of the image,
and processing them in a TBB parallel for loop. Each thread then traces
a packet of rays in SIMD, using Embree to traverse the ray packet, and ISPC
kernels to shade the packet's rays in parallel. The most similar production film renderer to OSPRay
is MoonRay~\cite{lee_vectorized_2017}, which uses larger ray streams instead of
SIMD-width packets to further improve memory coherence.

To efficiently distribute
rendering work and communicate across multiple nodes on a cluster, OSPRay leverages
a Distributed FrameBuffer~\cite{usher_scalable_2019} and MPI. Finally, to provide
a high-quality image at low sample counts OSPRay uses Intel's Open Image Denoise
library~\cite{oidn} for post-process denoising.

\paragraph*{ISPC}
ISPC is a compiler for a C-like language for writing single-program multiple data (SPMD)
kernels which are executed in SIMD on the CPU's vector lanes. The code is written
as a serial program which at runtime is executed in parallel, with a program instance
run per-CPU vector lane in a model roughly similar to GLSL, HLSL, CUDA, and OpenCL.
The group of program instances running on a vector unit is referred to as a ``gang''.
In contrast
to GPU programming languages, ISPC runs on the CPU in the same memory space as
the calling program, and can share pointers with the ``host'' program
or even call back into the host code. Directly sharing pointers with the rendering
kernels is especially valuable
for large scenes which already struggle to fit in memory, as this removes the need to
make a copy of the data to pass to the compute device.

ISPC's support for calling back into the host code directly is useful
for introducing vectorization into existing large
codebases and interfacing with non-vectorized code, without requiring
a complete re-write. As discussed in
Section~\ref{sec:ptex}, in this work we leveraged this capability to allow
our rendering and shading code written in ISPC to use the Ptex library
for texturing. Although the program gang must be serialized to call out
to the serial host code, this enables interoperability with existing
code for texturing, on demand model loading, etc., which would either be
difficult or impossible to port directly to ISPC.


Moreover, ISPC allows for easily writing portable vectorized code, which
is highly desirable when deploying renderers across a wide range of hardware. 
This portability is achieved by compiling a multi-target binary, which includes
specialized code paths for each backend supported by ISPC.
At runtime ISPC will then pick the correct code to run from this binary for the target architecture.
The portability and performance provided by ISPC have made it our language of choice
for implementing the core kernels of OSPRay, with higher-level scene setup, multi-threading,
and multi-node code implemented in C++.

Compared to alternatives for achieving vectorization on CPUs, e.g. compiler pragmas,
OpenCL, OpenMP, etc., we have found ISPC to provide better and more reliable performance.
A key drawback of auto-vectorization and compiler pragmas is that they
can easily break when control flow diverges, and revert to fully scalar code.
In contrast, ISPC is explicitly a SPMD on SIMD model and will still vectorize the code,
though at the cost of introducing control flow masking to handle possible divergence within
a gang.
Although diverging control flow in ISPC comes with a cost, it is far more desirable
to pay this cost and keep the code vectorized than to fall back
to completely scalar code in most cases.


However, some care must be taken when writing high-performance code in ISPC.
While a complete discussion of performance
considerations is beyond the scope of this paper\footnote{See the ISPC performance guide:
\url{http://ispc.github.io/perfguide.html}}, we discuss a few which are directly applicable
to the task of path tracing large, complex models.
To minimize control flow masking and allow greater use of scalar registers, we
recommended to use \texttt{uniform} variables wherever applicable.
A \texttt{uniform} variable in ISPC is one which is the same across
all vector lanes, and can be placed in a scalar register. Moreover, when a branch
depends on a \texttt{uniform} variable the control flow within a
gang is known to not diverge, allowing the compiler to avoid emitting
control flow masking instructions.

We also recommend to use ISPC in 32-bit addressing mode. In this mode, all pointers used
in ISPC kernels map to 32-bit offsets relative to a \texttt{uniform} 64-bit pointer.
This allows the compiler to use faster 32-bit address computations and
scatter/gather intrinsics, leading to significant performance gains.
However, when rendering large data sets 32-bit offsets may be
insufficient to access large data arrays of geometry or texture data. In such
cases we treat the single array as multiple subarrays, each indexable by
32-bit offsets from a 64-bit pointer, and use ISPC's \texttt{foreach\_unique}
execution construct to iterate over the unique subarrays being accessed by each
program instance.
We found that even on the Moana Island Scene, there were no objects
large enough to require emulating 64-bit addressing in this manner.

\paragraph*{The Distributed FrameBuffer}
To efficiently distribute rendering work among nodes and combine
the partial results produced by each node, OSPRay uses a
Distributed FrameBuffer~\cite{usher_scalable_2019}. The Distributed FrameBuffer (DFB)
is a general framework for executing image compositing and processing tasks
for distributed renderers through a distributed,
asynchronous tile processing pipeline.
Along with standard image- and data-distributed rendering,
the DFB supports more advanced configurations
where scene data can be partially replicated among nodes,
or some of the scene fully replicated and combined with distributed geometry.

A ``distributed renderer'' implemented using the DFB consists of a render,
responsible for producing image tiles, and a tile operation, which combines the
tiles received for some image tile into a single final tile.
A tile operation can be a simple averaging to combine multiple samples, or alpha-blended
depth-compositing, e.g., for data-distributed rendering.
After the tile operation is run,
additional post-processing tasks can be performed, e.g., tone-mapping.
The DFB distributes the execution of the tile operations and post-processing
tasks among the processes by assigning tile owners to run tasks for each tile
in round-robin order among the nodes.

To render the Moana Island Scene in parallel on multiple nodes we use
the image-parallel renderer in OSPRay.
The image-parallel renderer works similar to other master-worker rendering
architectures previously used in, e.g., Manta~\cite{bigler_design_2006}
and OpenRT~\cite{wald_flexible_2002}.
This renderer is exposed through OSPRay's \texttt{MPIOffloadDevice},
which transparently distributes the scene data to a set of worker
processes running on the compute nodes.
These workers then render the scene using the image-parallel distributed
renderer implemented with the DFB.
The rendering work is distributed
among the nodes by assigning the image tiles round-robin to each node
to provide even work distribution for most scenes.
Each node then renders its assigned tiles in parallel using multiple
threads. At the end of the frame the final tiles are gathered onto the head
node to display the image.

\begin{figure}
  \centering
  \adjustbox{trim={{.1\width} 0 {.4\width} 0},clip}{
	  \includegraphics[width=0.98\columnwidth]{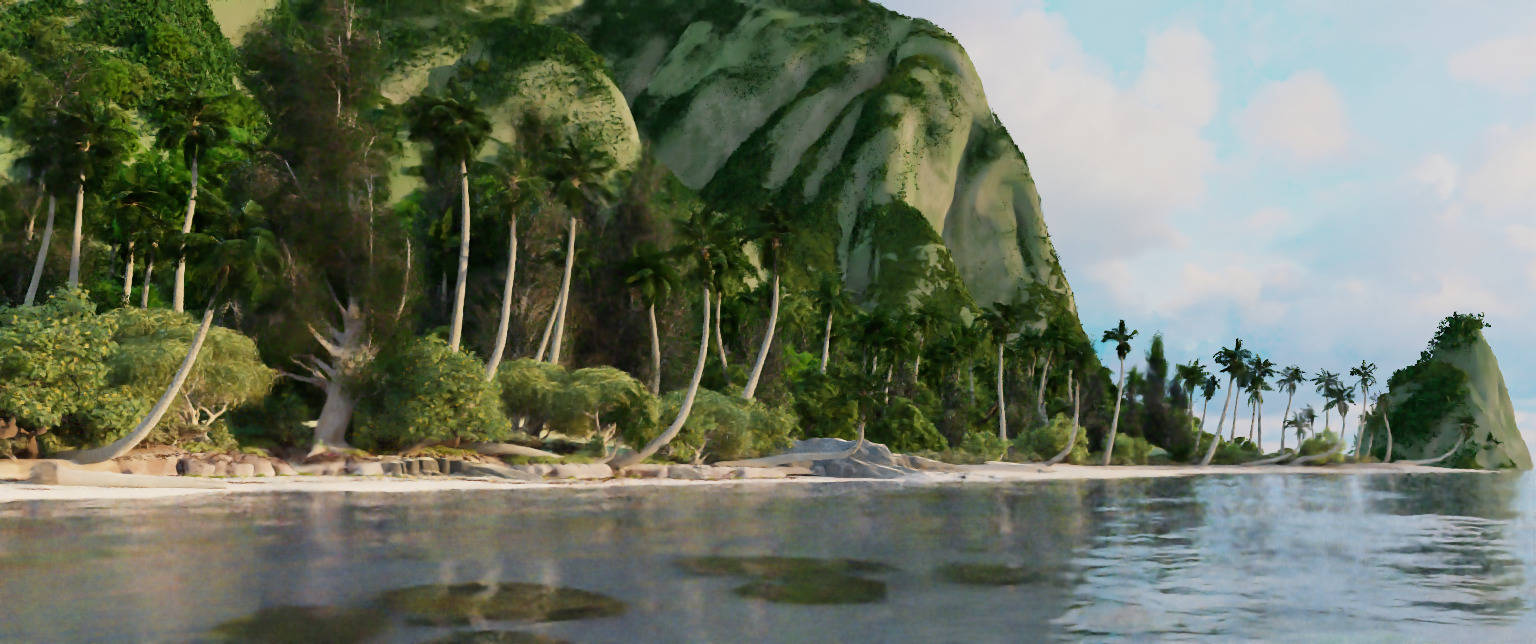}}
  \adjustbox{trim={{.1\width} 0 {.4\width} 0},clip}{
	  \includegraphics[width=0.98\columnwidth]{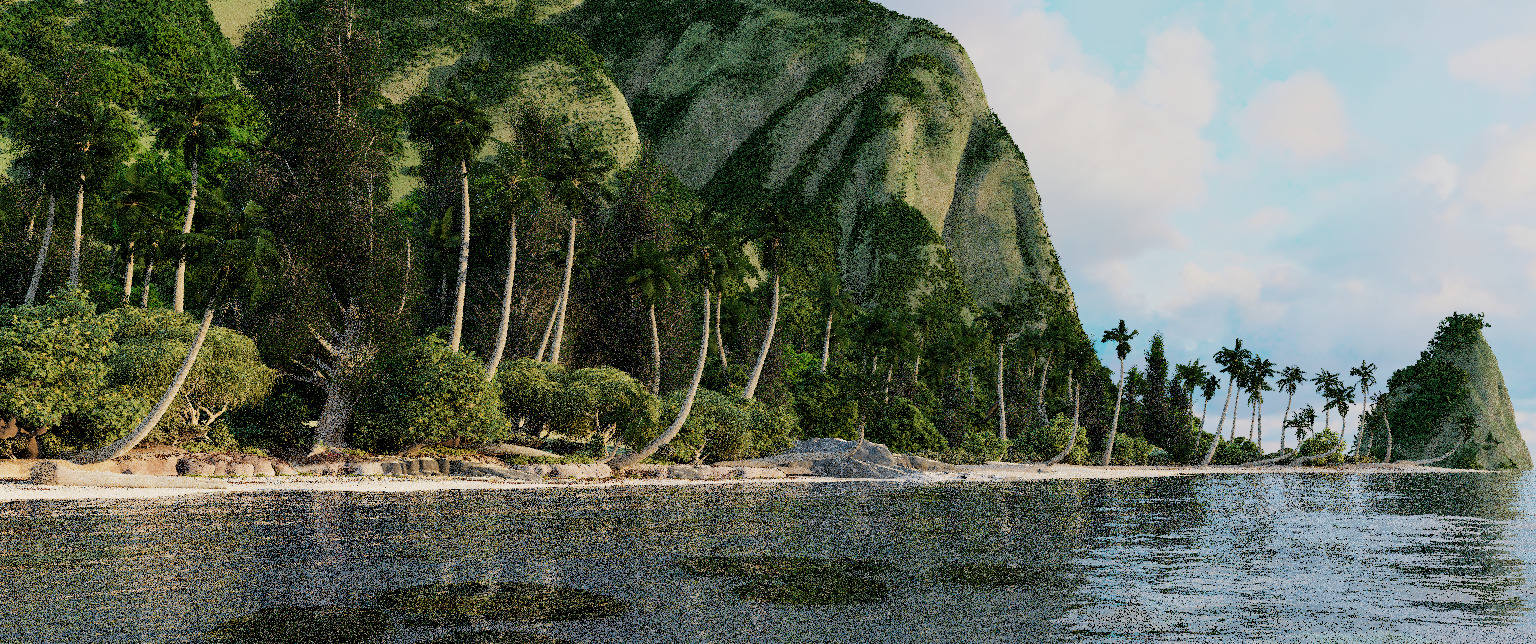}}
  \\
  \vspace*{-1em}
  \caption{\label{fig:denoising-1spp}%
  A crop of the Shot view, with and without denoising at one sample per-pixel.
	At the start of the progressive accumulation the denoiser provides a
	significant improvement in image quality, even at very low sample rates.}
\end{figure}

\begin{figure*}
  \centering
	\includegraphics[width=0.32\textwidth]{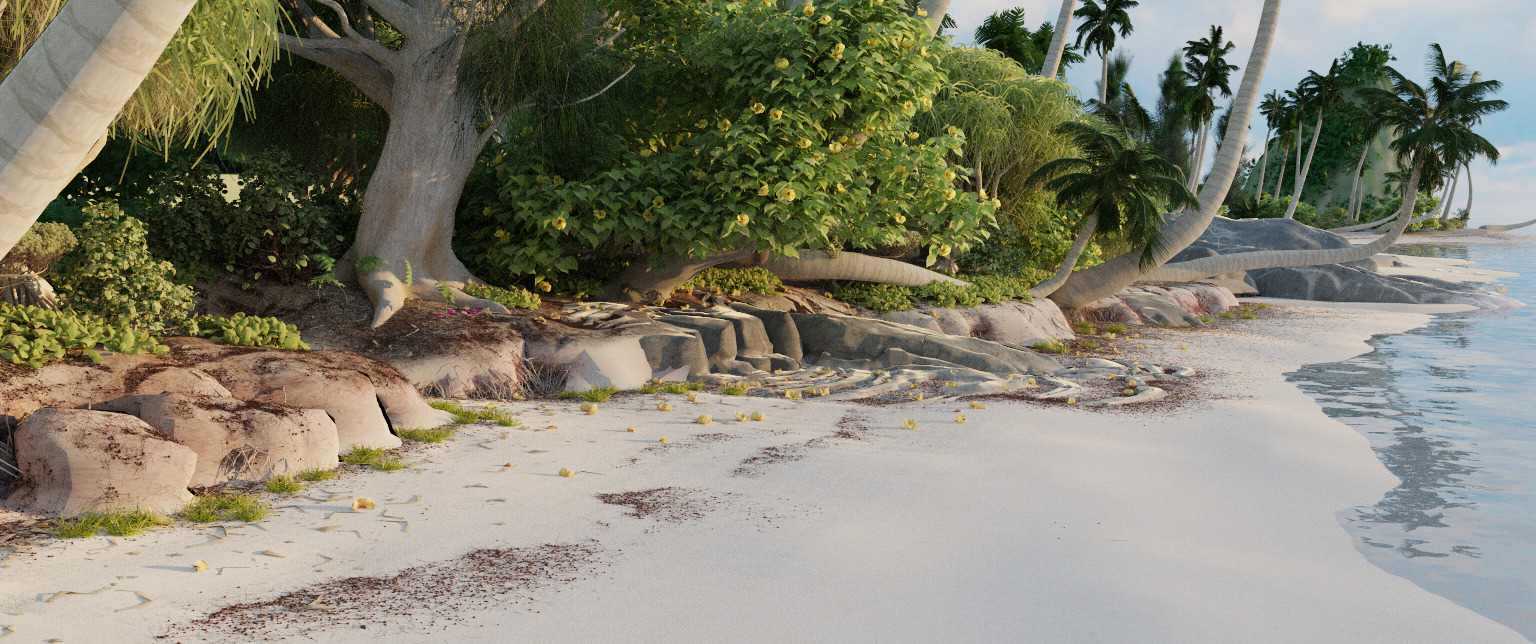}
	\includegraphics[width=0.32\textwidth]{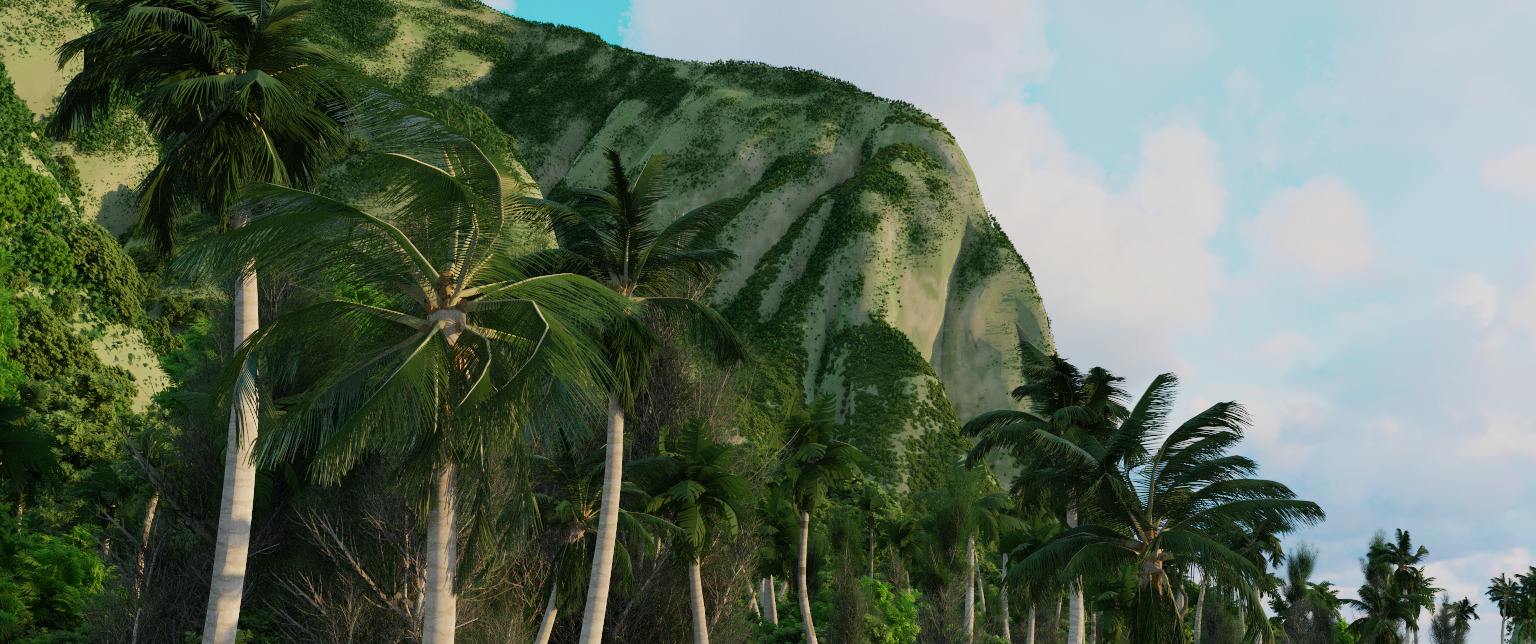}
	\includegraphics[width=0.32\textwidth]{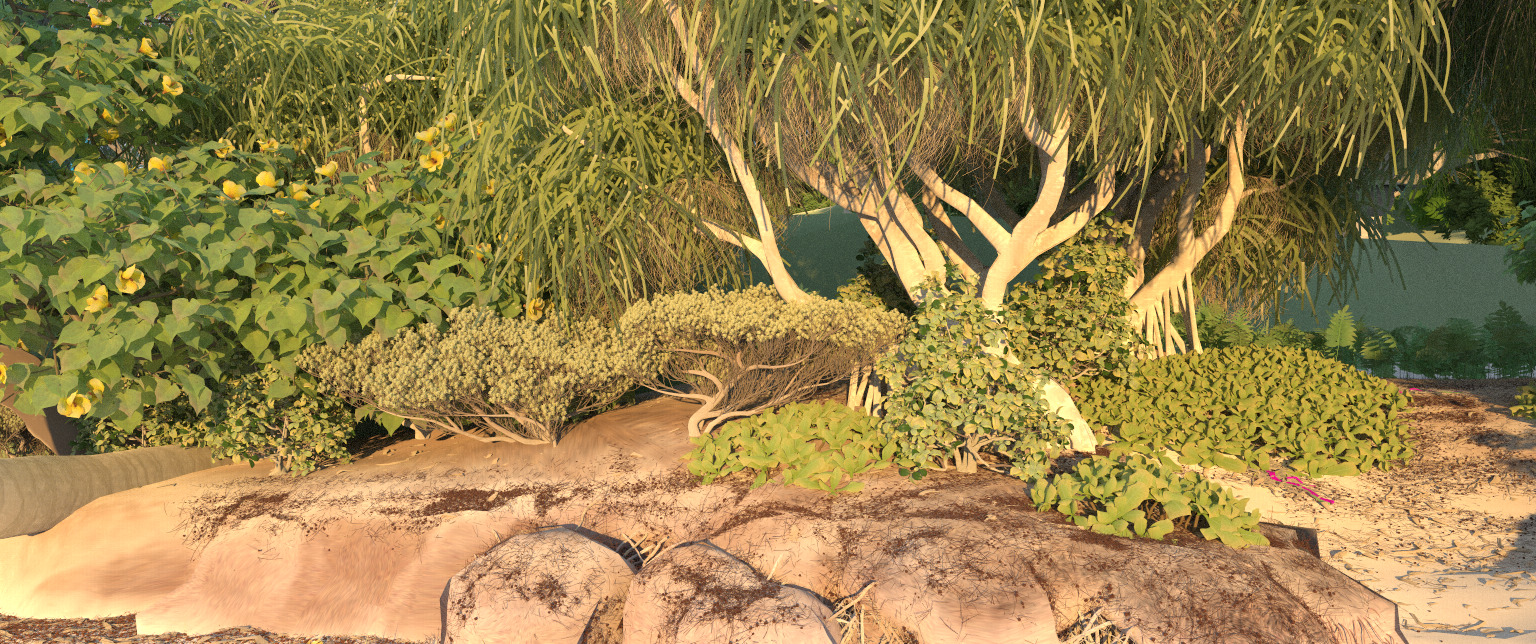}
        \vspace{-1em}
	\caption{\label{fig:reference-frames}%
	The camera viewpoints used in our benchmarks. The Beach
	camera (left) contains a mix of geometries filling roughly the entire
	image and a range of materials, and averages 252ms per-frame.
	The Palms camera (center) consists of
	diffuse materials, with a mix of large and sub-pixel geometry, and
	averages 183ms per-frame.
	The Dunes camera (right) consists of diffuse materials with a range of geometry covering
	the entire image, and averages 339ms per-frame.}
\end{figure*}

\paragraph*{Denoising}



Beyond increasing compute power and improved efficiency,
arguably one of the biggest breakthroughs in recent years
for interactive path tracing performance was the introduction
of denoising techniques~\cite{bitterli_denoise_2016,mara_denoise_2017,schied_denoise_2017,
nv-denoise,disney-denoise}.
Recent techniques based on machine learning have been shown
to be fast and capable of high quality images with few samples per-pixel,
see e.g., Chaitanya et al.~\shortcite{nv-denoise} or Bako et al.~\shortcite{disney-denoise}.
However, current approaches are not without limitations when
applied to interactive rendering of production content.
Real-time denoising techniques (e.g.,~\cite{schied_denoise_2017,mara_denoise_2017,nv-denoise})
can provide smooth images with just a sample per-pixel, but are 
too approximate for production; while denoising techniques
for production rendering (e.g.,~\cite{bitterli_denoise_2016,disney-denoise})
can provide better quality, they require a higher initial sampling
rate, more image features, and do not run in real-time.


In this work we use Intel's Open Image Denoise library~\cite{oidn},
which is a fast CPU implementation of
a denoiser in the spirit of Chaitanya et al.~\shortcite{nv-denoise}.
To preserve as much image detail as possible we provide not only color but albedo
and normal buffers as well to the denoiser.
While at very low sample rates we do observe visible blurring artifacts,
in particular on trees, bushes, and reflections, these
are preferable to the unfiltered image (see~Figure~\ref{fig:denoising-1spp}).
As additional samples are accumulated over time and the image converges,
the better image quality provided to the denoiser results in better
handling of these fine detail features.


When run in parallel on multiple nodes the denoising is performed on the
head node after the image is rendered. This allows the denoiser
to access the entire image at once, but clearly poses a scalablity
issue for large images or expensive denoisers. To avoid the workers
remaining idle while the head
node is denoising the frame, we run the denoising in parallel to the rendering.
When a frame is finished we begin denoising it on the
head node and immediately start rendering the next frame on the workers.
Distributing the denoising work to be run in parallel across the workers
through the DFB would improve performance for large images and expensive
denoisers, though would require the workers to perform some form of neighboring
pixel exchange to provide the required data for the denoiser.
How this exchange can be done efficiently remains an
interesting follow-on effort.


\subsection{USD Moana}
\begin{figure}
  \centering
  \includegraphics[width=0.48\textwidth]{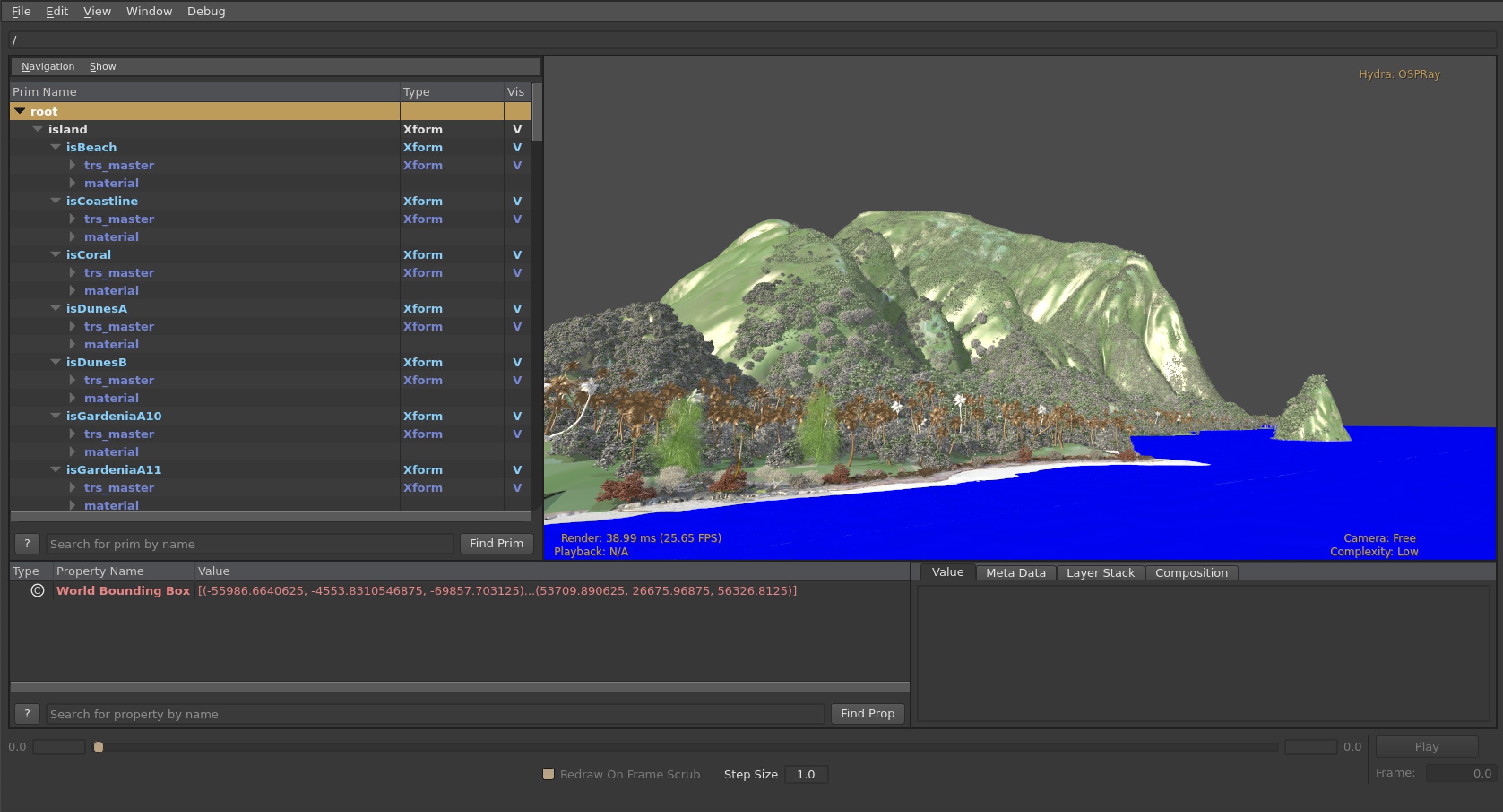}
  \\
  \vspace*{-1em}
	\caption{\label{fig:usd}%
    A rendering of the work in progress USD Moana Island Scene
    using HdOSPRay.}  
\end{figure}

\begin{figure}
  \centering
	\adjustbox{trim={{.2\width} 0 {.3\width} 0},clip}{
		\includegraphics[width=0.98\columnwidth]{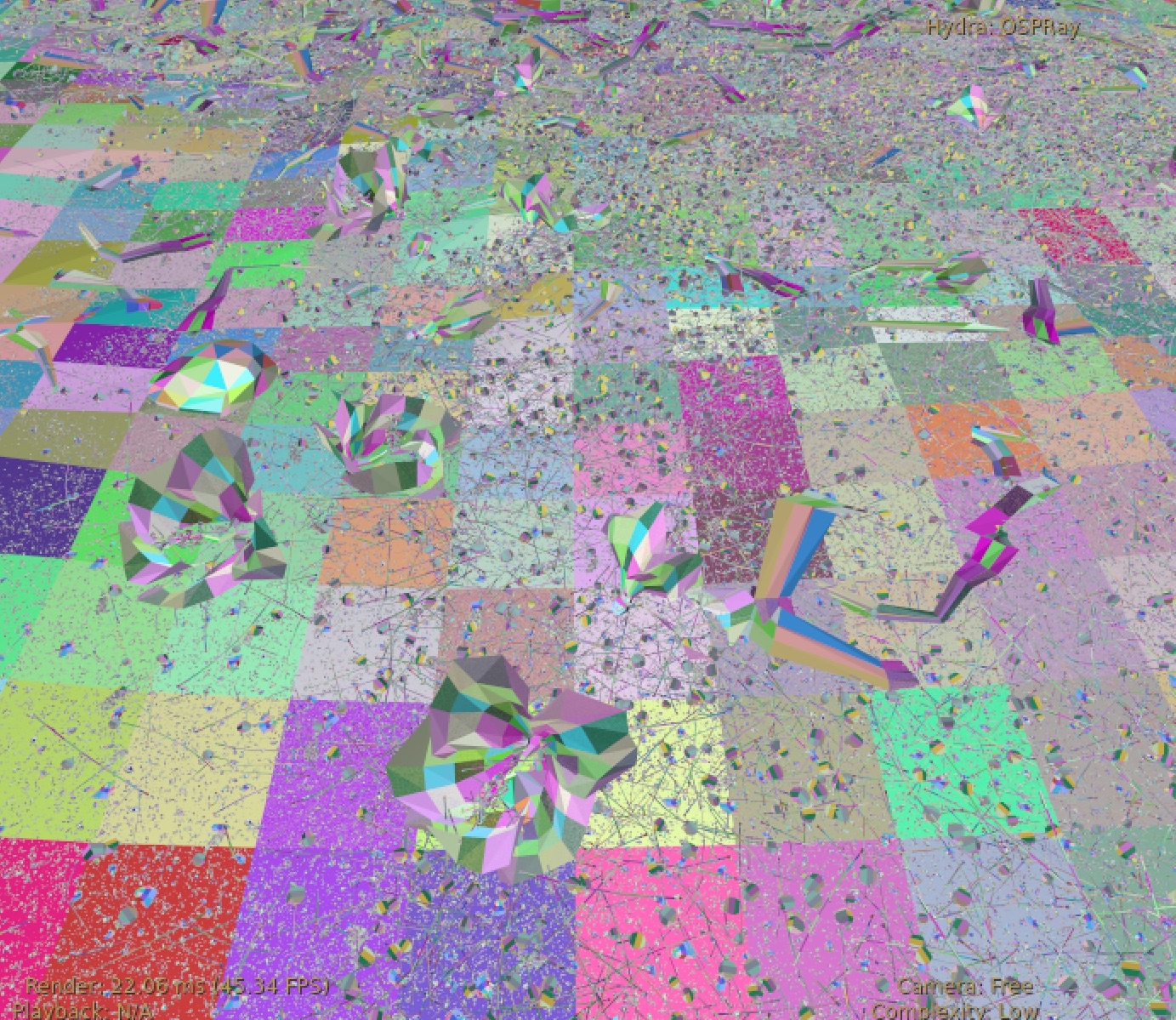}}
	\adjustbox{trim={{.2\width} 0 {.3\width} 0},clip}{
		\includegraphics[width=0.98\columnwidth]{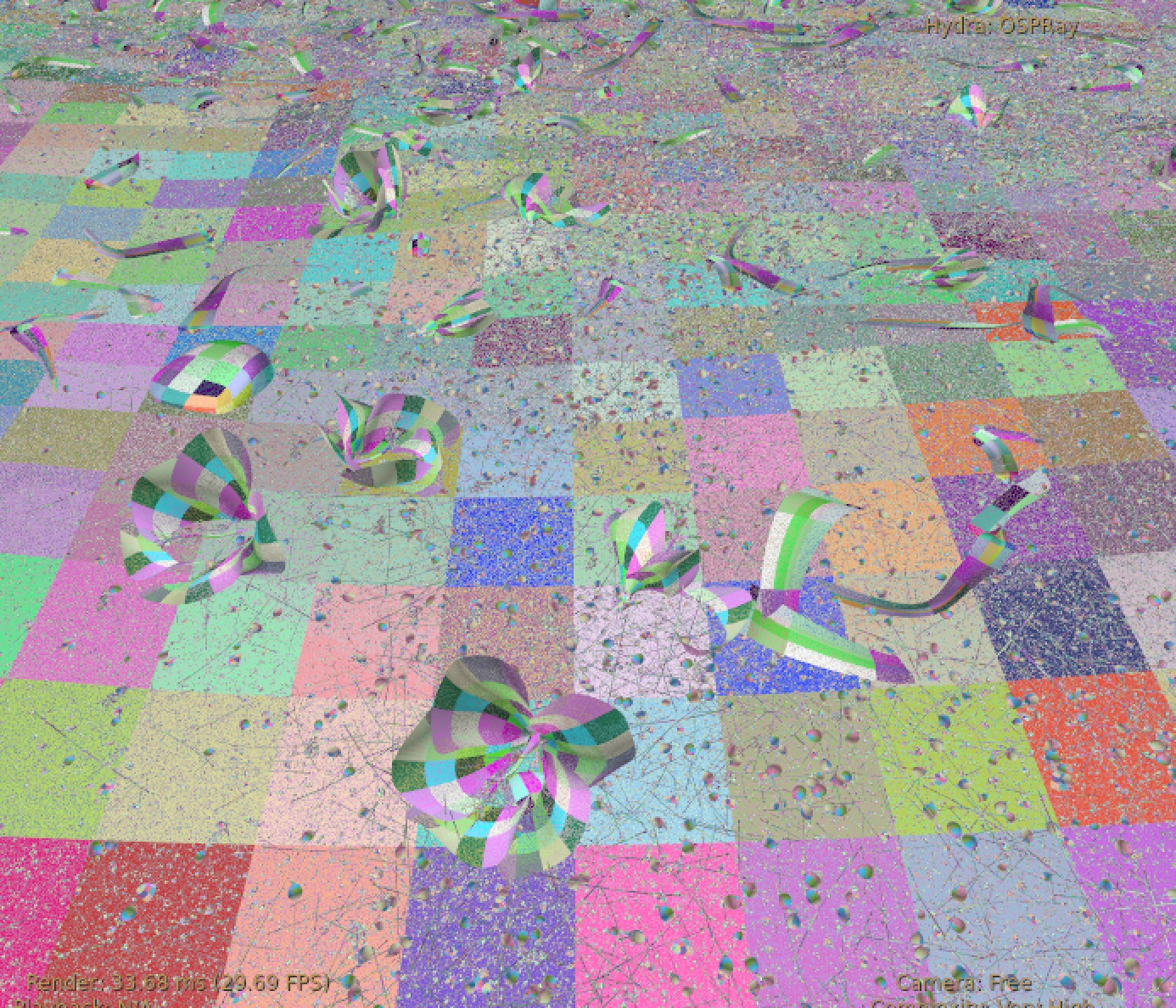}}
  \\
  \vspace*{-1em}
	\caption{\label{fig:subd}%
    Renderings of flowers using HdOSPRay and subdivision surfaces with 
		tessellation rates of 0, left, and 8, right.}  
\end{figure}

Pixar's Universal Scene Description format, USD, was designed for 
fast loading, rendering, and collaborative editing of large-scale 
production assets~\cite{poh_2018} which makes it a promising
standardized alternative to using PBRT or our custom binary format.
  Originally an internal scene graph format
used by Pixar in production, USD was recently released as an open source project,
 and has subsequently seen 
a wide adoption across other studios and even use within the game 
industry~\cite{Blevins_2018}.
  Due to the lack of publicly available USD datasets
for the wider community to develop and test new techniques with,
Disney began work on an additional version of the Moana Island Scene converted to USD.
As of this writing, the USD version of the dataset remains a work in progress,
and does not yet match the full scale or correctness of the 
publicly released JSON or PBRT versions of the model.
It does, however, present a widely used file format that exhibits
significantly improved loading times over pbrt and supports subdivision surfaces 
which were missing in the pbrt conversion.  
We loaded and rendered the USD Moana Island Scene using the OSPRay
backend of the Hydra rendering layer in USD,
HdOSPRay\footnote{\url{https://github.com/ospray/hdospray}} (see~Figure~\ref{fig:usd}).
The lack of detail missing from
only using the subdivision cages, a refinement level of 0,
compared to using a tessellation rate of 8 can be seen in 
Fig.~\ref{fig:subd}.

\section{Results}
\label{sec:results}

We evaluate the performance of our renderer
using the predefined Shot, Beach, Palms, and Dunes camera positions provided
in the PBRT scene file.
The camera positions chosen cover various configurations in terms
of the directly visible geometries and materials, covering a range
of cost per-pixel (see~Figures~\ref{fig:teaser} and \ref{fig:reference-frames})

The benchmarks are rendered at the film aspect ratio at a resolution
of $1536\times644$ and a maximum path depth of five.
We use Ptex's caching system to manage loading textures, and configure it
to allow for an unlimited amount of cache memory and 100 open files.
The Ptex cache does take some period to warm up, we found that the first 25 to 30
frames take much longer than subsequent frames, with the first few
frames taking orders of magnitude longer as frequently accessed
textures are loaded into the cache.
To benchmark the renderer after this warm up period
we use the first 64 frames as warm up frames, and
measure performance over the next 64.
Our benchmarks are run using nine Intel Skylake Xeon nodes
on the Texas Advanced Computing Center's Stampede2 system, with one
head node and eight worker nodes. Each
node has two Intel Xeon Platinum 8160 processors and 192GB of DDR4 RAM.

%
%
%
%
%
%
%
%
%
%

\begin{figure}
	\centering
	\adjustbox{trim={{.2\width} 0 {.3\width} 0},clip}{
		\includegraphics[width=0.98\columnwidth]{png/palms-cam-beauty.jpg}}
	\adjustbox{trim={{.2\width} 0 {.3\width} 0},clip}{
		\includegraphics[width=0.98\columnwidth]{png/dunes-cam-beauty.jpg}}
	\\
	\adjustbox{trim={{.2\width} 0 {.3\width} 0},clip}{
		\includegraphics[width=0.98\columnwidth]{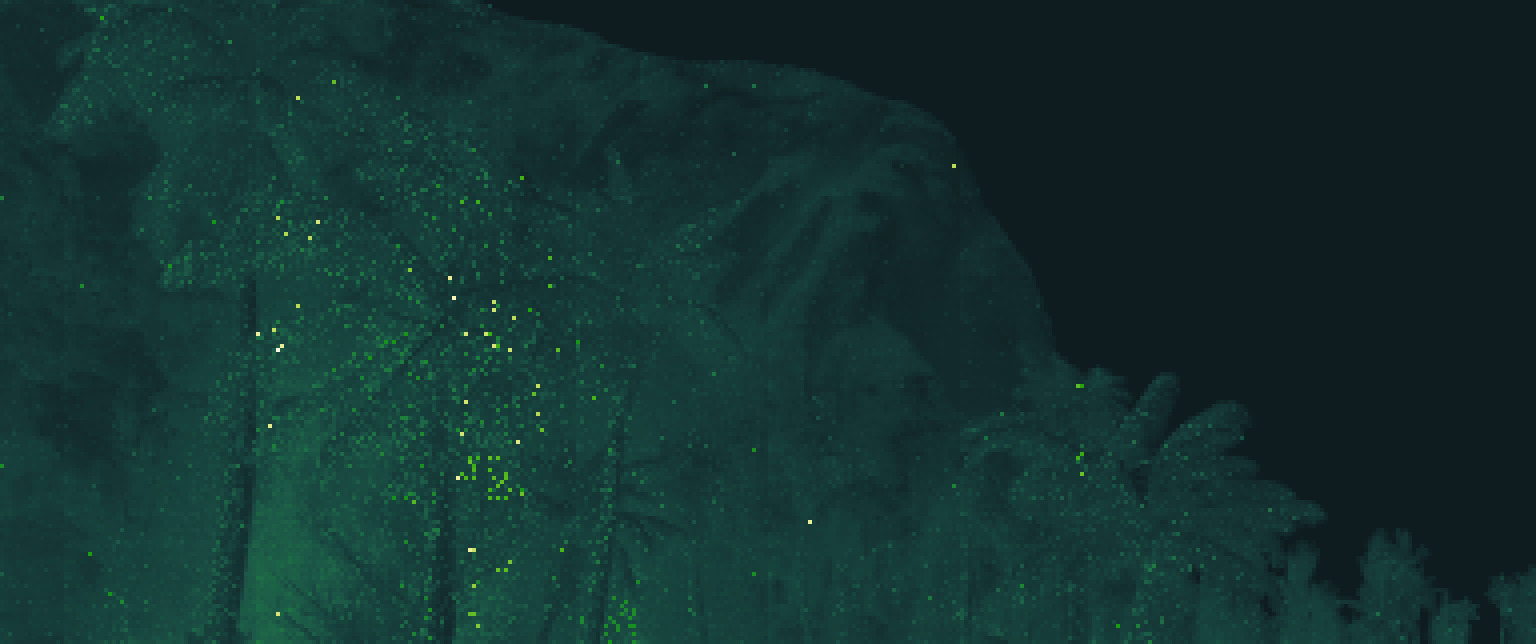}}
	\adjustbox{trim={{.2\width} 0 {.3\width} 0},clip}{
	  \includegraphics[width=0.98\columnwidth]{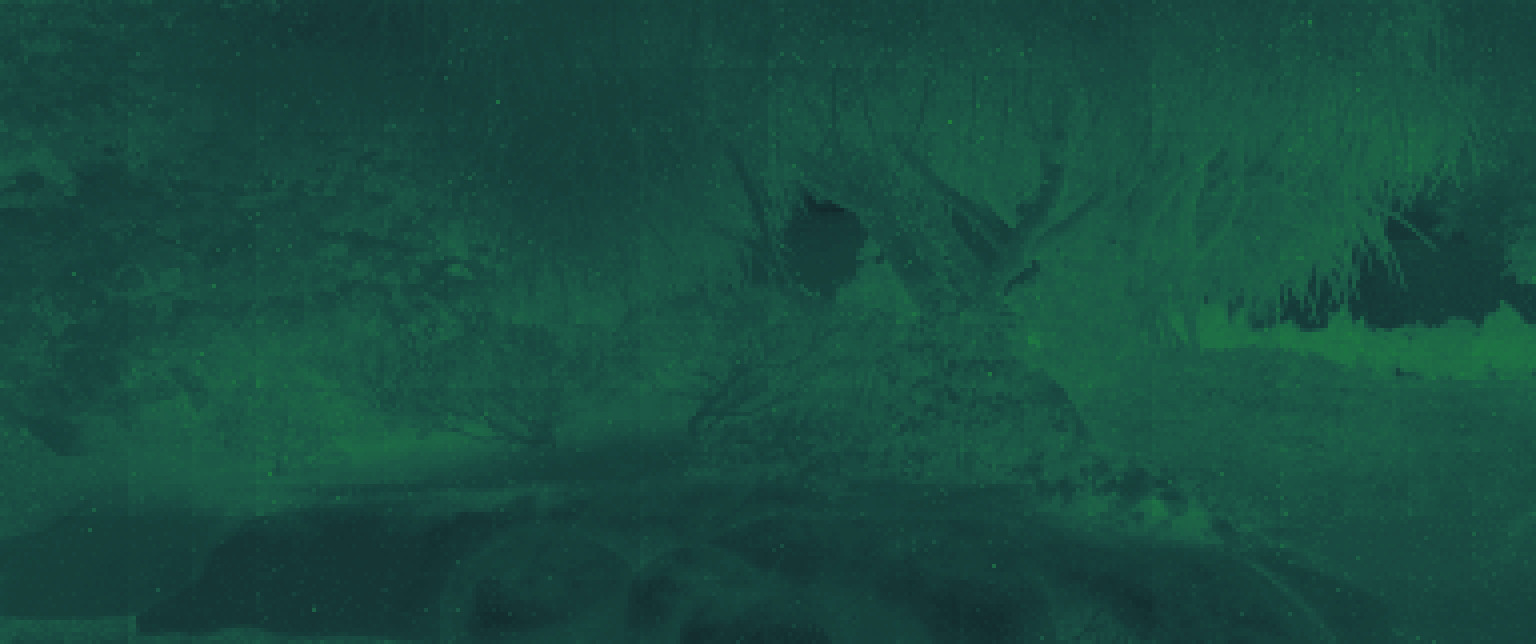}}
  \\
  \vspace*{-1em}
	\caption{\label{fig:heat-map}%
	Crops of two exemplary views on the cost-per-pixel spectrum.
	Left: The Palms view is cheap on average with some hot spots, and achieves 5.46~FPS and
	39.38 Mray/s.
	Right: The Dunes view is far more expensive on average, and achieves
	2.95~FPS at 36.32 Mray/s.
	Bottom: The same, colored by cost per-pixel from dark (low) to light (high).
	Both images use the same heat-map scale.}
\end{figure}

\begin{table}
	\centering
	\caption{\label{tab:profile}%
	Compute time breakdowns for each of the benchmarked views.
	We find that the majority of time is spent in Embree performing ray
	traversal and intersection tests, with sampling and shading BRDFs
	taking the bulk of the remaining time. Surprisingly little time
	is spent in Ptex, which is likely attributable to the texture caching
	performed by the library.}
	\begin{tabular}{lrrrr}
		\toprule
		Component & Shot & Beach & Palms & Dunes \\
		\midrule
		Embree (T. \& I.)   & 67.28\% & 70.17\% & 71.20\% & 74.99\% \\
		PostIsec            & 7.10\%  & 5.96\%  & 6.99\%  & 5.13\% \\
		Ptex                & 2.26\%  & 2.89\%  & 1.25\%  & 2.27\% \\
		Sample \& Shade     & 22.74\% & 20.45\% & 19.87\% & 17.21\% \\
		Other               & 0.62\%  & 0.53\%  & 0.69\%  & 0.39\%   \\
		\bottomrule
\end{tabular}
\end{table}

In terms of overall rendering performance, we find that eight worker nodes are
sufficient to provide interactive rendering. For each
camera position, we measured the average ray tracing time to be:
207ms for Shot (36.95~Mray/s),
252ms for Beach (35.29~Mray/s),
183ms for Palms (39.38~Mray/s),
and 339ms for Dunes (36.32~Mray/s).
The image denoising cost depends only on the
number of pixels being processed, and takes on average 130ms per-frame
across all the benchmarked views.

The ray tracing times correspond to larger differences
in the number of rays actually processed and shaded per-frame, due to the differences
in the scenes being rendered.
In the Shot and Palms views a large portion of the image only sees the background,
and in the Shot view additional camera rays reflect off the
water surface into the background. In contrast, the Beach
and Dunes views are largely filled with dense geometry with smooth materials,
resulting in a large number of diffuse bounces and thus rays traced
(also see Figure~\ref{fig:heat-map}). For example, the average number of rays traced
per-pixel on the Palms scene is just 7.29, while the Dunes traces 12.45 per-pixel,
corresponding to the lowest and highest average rays per-pixel across
the benchmarks, respectively. 

To determine where time is spent within the renderer, we break down
the total compute time for each view by component in Table~\ref{tab:profile}.
Across all scenes we observe that the majority of time (67-75\%) is spent
in Embree, tracing rays and intersecting geometry, with the second largest
amount of time spent sampling and shading BRDFs. The material model
currently used in OSPRay's path tracer returns a set of BRDFs from the
material; and with potentially different materials hit by each ray in a packet
and different sets of BRDFs returned by these materials, the shading code
can become quite expensive and nearly serialized for a packet.
The PostIsec time measures the time spent
computing the properties needed to shade the BRDF, namely the surface normals,
texture coordinates, and so on.

The result we found the most surprising was how little time was spent
in Ptex after the warm up period. After the bulk of texture data
which is needed for the scene has been loaded into the
cache, the time spent sampling textures drops significantly. Even in the
Beach view, where a large portion of the scene is visible, Ptex lookups only
account for 2.89\% of the total compute time for a frame. We do
note that this is not the case during the warm up frames, especially
when data is first being read from disk and cached. During the warm up
period the Ptex and Sample \& Shade components together take up the bulk of compute
time, up to 75\% in some cases, as required texture data is fetched and cached.


\section{Discussion}
\label{sec:discussion}
In this paper we have presented the challenges encountered and solutions
developed to achieve interactive rendering performance on the Moana
Island Scene. Such production scenes present a significant challenge
to interactive rendering, from loading the data at all, to rendering
it interactively at full quality. As production scenes
of this scale are typically not available to the broader research
community, we hope that by presenting our experiences and difficulties
in working with this asset, this paper can provide guidance to other researchers
beginning to work with the Moana Island Scene or other similar production assets.

With tools and renderers capable of interactive rendering at full geometric and shading quality
on production scenes, artists will be able to iterate more quickly
on modeling and design of film assets. To this end, we are working on
integrating the rendering system presented with production tools, and developing
native support for USD through HdOSPRay.
As modeling tools for film become more interactive, it is interesting
to consider whether this faster feedback loop between artist changes
and results will change the underlying assets, or the films themselves.
Given the ability to truly explore scenes interactively, directors may
frame shots differently, or artists be better able
to adjust lighting and materials to achieve the desired look.



Finally, while film resolutions are increasing, the rate at which geometric
and texture complexity is increasing far outpaces it. When the
renderer is only able to parallelize the rendering work over the pixels
and samples in an image,
there is an inherent limit on the amount of parallelism which can be extracted.
To this end, it may be valuable to consider
data-distributed rendering of such assets, where subregions of the data
are assigned to different nodes, and rays or data moved as needed during
rendering. A data-distributed approach may also allow for GPU-based
interactive rendering of such production scenes, where memory is more
constrained than on a CPU.

\begin{acks}
    The authors would like to thank Disney, and in particular Rasmus Tamstorf,
    for making the Moana Island Scene publicly available, for granting access
    to early versions and for assistance with the data.
    The authors thank the Texas Advanced Computing Center (TACC) at
    the University of Texas at Austin for providing HPC resources that
    have contributed to the results reported in this paper.
\end{acks}

\bibliographystyle{ACM-Reference-Format}
\bibliography{moana}


\begin{thebibliography}{36}


\ifx \showCODEN    \undefined \def \showCODEN     #1{\unskip}     \fi
\ifx \showDOI      \undefined \def \showDOI       #1{#1}\fi
\ifx \showISBNx    \undefined \def \showISBNx     #1{\unskip}     \fi
\ifx \showISBNxiii \undefined \def \showISBNxiii  #1{\unskip}     \fi
\ifx \showISSN     \undefined \def \showISSN      #1{\unskip}     \fi
\ifx \showLCCN     \undefined \def \showLCCN      #1{\unskip}     \fi
\ifx \shownote     \undefined \def \shownote      #1{#1}          \fi
\ifx \showarticletitle \undefined \def \showarticletitle #1{#1}   \fi
\ifx \showURL      \undefined \def \showURL       {\relax}        \fi
\providecommand\bibfield[2]{#2}
\providecommand\bibinfo[2]{#2}
\providecommand\natexlab[1]{#1}
\providecommand\showeprint[2][]{arXiv:#2}

\bibitem[\protect\citeauthoryear{{4A Games}}{{4A Games}}{2019}]%
        {metro}
\bibfield{author}{\bibinfo{person}{{4A Games}}.}
  \bibinfo{year}{2019}\natexlab{}.
\newblock \bibinfo{title}{Metro {Exodus}}.
\newblock
\newblock


\bibitem[\protect\citeauthoryear{Bako, Vogels, McWilliams, Meyer, Novak,
  Harvill, Sen, DeRose, and Rousselle}{Bako et~al\mbox{.}}{2017}]%
        {disney-denoise}
\bibfield{author}{\bibinfo{person}{Steve Bako}, \bibinfo{person}{Thijs Vogels},
  \bibinfo{person}{Brian McWilliams}, \bibinfo{person}{Mark Meyer},
  \bibinfo{person}{Jan Novak}, \bibinfo{person}{Alex Harvill},
  \bibinfo{person}{Pradeep Sen}, \bibinfo{person}{Tony DeRose}, {and}
  \bibinfo{person}{Fabrice Rousselle}.} \bibinfo{year}{2017}\natexlab{}.
\newblock \showarticletitle{"Kernel-Predicting Convolutional Networks for
  Denoising Monte Carlo Renderings"}.
\newblock \bibinfo{journal}{\emph{ACM Transactions on Graphics (Proceedings of
  SIGGRAPH 2017)}} \bibinfo{volume}{36}, \bibinfo{number}{4}
  (\bibinfo{year}{2017}).
\newblock


\bibitem[\protect\citeauthoryear{Bala}{Bala}{2018}]%
        {tog-special-issue}
\bibfield{editor}{\bibinfo{person}{Kavita Bala}} (Ed.).
  \bibinfo{year}{2018}\natexlab{}.
\newblock \showarticletitle{Special Issue On Production Rendering and Regular
  Papers}.
\newblock \bibinfo{journal}{\emph{ACM Transactions on Graphics}}
  \bibinfo{volume}{37}, \bibinfo{number}{3} (\bibinfo{year}{2018}).
\newblock


\bibitem[\protect\citeauthoryear{Bigler, Stephens, and Parker}{Bigler
  et~al\mbox{.}}{2006}]%
        {bigler_design_2006}
\bibfield{author}{\bibinfo{person}{James Bigler}, \bibinfo{person}{Abe
  Stephens}, {and} \bibinfo{person}{Steven~G. Parker}.}
  \bibinfo{year}{2006}\natexlab{}.
\newblock \showarticletitle{Design for {{Parallel Interactive Ray Tracing
  Systems}}}. In \bibinfo{booktitle}{\emph{2006 {{IEEE Symposium}} on
  {{Interactive Ray Tracing}}}}.
\newblock


\bibitem[\protect\citeauthoryear{Bitterli, Rousselle, Moon, Iglesias-Guitián,
  Adler, Mitchell, Jarosz, and Novák}{Bitterli et~al\mbox{.}}{2016}]%
        {bitterli_denoise_2016}
\bibfield{author}{\bibinfo{person}{Benedikt Bitterli}, \bibinfo{person}{Fabrice
  Rousselle}, \bibinfo{person}{Bochang Moon}, \bibinfo{person}{José~A.
  Iglesias-Guitián}, \bibinfo{person}{David Adler}, \bibinfo{person}{Kenny
  Mitchell}, \bibinfo{person}{Wojciech Jarosz}, {and} \bibinfo{person}{Jan
  Novák}.} \bibinfo{year}{2016}\natexlab{}.
\newblock \showarticletitle{Nonlinearly Weighted First-order Regression for
  Denoising Monte Carlo Renderings}.
\newblock \bibinfo{journal}{\emph{Computer Graphics Forum (Proceedings of
  EGSR)}} \bibinfo{volume}{35}, \bibinfo{number}{4} (\bibinfo{year}{2016}).
\newblock


\bibitem[\protect\citeauthoryear{Blevins and Murray}{Blevins and
  Murray}{2018}]%
        {Blevins_2018}
\bibfield{author}{\bibinfo{person}{Alan Blevins} {and} \bibinfo{person}{Mike
  Murray}.} \bibinfo{year}{2018}\natexlab{}.
\newblock \showarticletitle{Zero to USD in 80 Days: Transitioning Feature
  Production to Universal Scene Description at Dreamworks}. In
  \bibinfo{booktitle}{\emph{ACM SIGGRAPH 2018 Talks}}
  \emph{(\bibinfo{series}{SIGGRAPH `18})}. \bibinfo{publisher}{ACM},
  \bibinfo{address}{New York, NY, USA}, Article \bibinfo{articleno}{53},
  \bibinfo{numpages}{2}~pages.
\newblock
\showISBNx{978-1-4503-5820-0}


\bibitem[\protect\citeauthoryear{Burley}{Burley}{2012}]%
        {disney-principled}
\bibfield{author}{\bibinfo{person}{Brent Burley}.}
  \bibinfo{year}{2012}\natexlab{}.
\newblock \showarticletitle{Physically-based {Shading} at {Disney}}.
\newblock In \bibinfo{booktitle}{\emph{SIGGRAPH 2012 Course Notes ``Practical
  Physically Based Shading in Film and Game Production''}}.
\newblock


\bibitem[\protect\citeauthoryear{Burley}{Burley}{2015}]%
        {disney-principled-bsdf}
\bibfield{author}{\bibinfo{person}{Brent Burley}.}
  \bibinfo{year}{2015}\natexlab{}.
\newblock \showarticletitle{Extending the {Disney} {BRDF} to a {BSDF} with
  Integrated Subsurface Scattering}.
\newblock In \bibinfo{booktitle}{\emph{SIGGRAPH 2015 Course Notes "Physically
  Based Shading in Theory and Practice "}}.
\newblock


\bibitem[\protect\citeauthoryear{Burley, Adler, Chiang, Driskill, Habel, Kelly,
  Kutz, Li, and Teece}{Burley et~al\mbox{.}}{2018}]%
        {hyperion}
\bibfield{author}{\bibinfo{person}{Brent Burley}, \bibinfo{person}{David
  Adler}, \bibinfo{person}{Matt Jen-Yuan Chiang}, \bibinfo{person}{Hank
  Driskill}, \bibinfo{person}{Ralf Habel}, \bibinfo{person}{Patrick Kelly},
  \bibinfo{person}{Peter Kutz}, \bibinfo{person}{Yining~Karl Li}, {and}
  \bibinfo{person}{Daniel Teece}.} \bibinfo{year}{2018}\natexlab{}.
\newblock \showarticletitle{The Design and Evolution of Disney's Hyperion
  Renderer}.
\newblock \bibinfo{journal}{\emph{ACM Transactions on Graphics}}
  \bibinfo{volume}{37}, \bibinfo{number}{3}, Article \bibinfo{articleno}{33}
  (\bibinfo{date}{July} \bibinfo{year}{2018}), \bibinfo{numpages}{22}~pages.
\newblock
\showISSN{0730-0301}


\bibitem[\protect\citeauthoryear{Burley and Lacewell}{Burley and
  Lacewell}{2008}]%
        {ptex}
\bibfield{author}{\bibinfo{person}{Brent Burley} {and} \bibinfo{person}{Dylan
  Lacewell}.} \bibinfo{year}{2008}\natexlab{}.
\newblock \showarticletitle{Ptex: Per-Face Texture Mapping for Production
  Rendering}.
\newblock \bibinfo{journal}{\emph{Computer Graphics Forum}}
  \bibinfo{volume}{27}, \bibinfo{number}{4} (\bibinfo{year}{2008}),
  \bibinfo{pages}{1155--1164}.
\newblock


\bibitem[\protect\citeauthoryear{Chaitanya, Kaplanyan, Schied, Salvi, Lefohn,
  Nowrouzezahrai, and Aila}{Chaitanya et~al\mbox{.}}{2017}]%
        {nv-denoise}
\bibfield{author}{\bibinfo{person}{Chakravarty R.~Alla Chaitanya},
  \bibinfo{person}{Anton Kaplanyan}, \bibinfo{person}{Christoph Schied},
  \bibinfo{person}{Marco Salvi}, \bibinfo{person}{Aaron Lefohn},
  \bibinfo{person}{Derek Nowrouzezahrai}, {and} \bibinfo{person}{Timo Aila}.}
  \bibinfo{year}{2017}\natexlab{}.
\newblock \showarticletitle{"Interactive Reconstruction of Monte Carlo Image
  Sequences using a Recurrent Denoising Autoencoder"}.
\newblock \bibinfo{journal}{\emph{ACM Transactions on Graphics (Proceedings of
  SIGGRAPH 2017)}} \bibinfo{volume}{36}, \bibinfo{number}{4}
  (\bibinfo{year}{2017}).
\newblock


\bibitem[\protect\citeauthoryear{Christensen, Fong, Shade, Wooten, Schubert,
  Kensler, Friedman, Kilpatrick, Ramshaw, Bannister, Rayner, Brouillat, and
  Liani}{Christensen et~al\mbox{.}}{2018}]%
        {renderman}
\bibfield{author}{\bibinfo{person}{Per Christensen}, \bibinfo{person}{Julian
  Fong}, \bibinfo{person}{Jonathan Shade}, \bibinfo{person}{Wayne Wooten},
  \bibinfo{person}{Brenden Schubert}, \bibinfo{person}{Andrew Kensler},
  \bibinfo{person}{Stephen Friedman}, \bibinfo{person}{Charlie Kilpatrick},
  \bibinfo{person}{Cliff Ramshaw}, \bibinfo{person}{Marc Bannister},
  \bibinfo{person}{Brenton Rayner}, \bibinfo{person}{Jonathan Brouillat}, {and}
  \bibinfo{person}{Max Liani}.} \bibinfo{year}{2018}\natexlab{}.
\newblock \showarticletitle{RenderMan: An Advanced Path-Tracing Architecture
  for Movie Rendering}.
\newblock \bibinfo{journal}{\emph{ACM Transactions on Graphics}}
  \bibinfo{volume}{37}, \bibinfo{number}{3}, Article \bibinfo{articleno}{30}
  (\bibinfo{date}{Aug.} \bibinfo{year}{2018}), \bibinfo{numpages}{21}~pages.
\newblock
\showISSN{0730-0301}


\bibitem[\protect\citeauthoryear{Cook, Carpenter, and Catmull}{Cook
  et~al\mbox{.}}{1987}]%
        {cook_reyes_1987}
\bibfield{author}{\bibinfo{person}{Robert~L. Cook}, \bibinfo{person}{Loren
  Carpenter}, {and} \bibinfo{person}{Edwin Catmull}.}
  \bibinfo{year}{1987}\natexlab{}.
\newblock \showarticletitle{The {Reyes} {Image} {Rendering} {Architecture}}. In
  \bibinfo{booktitle}{\emph{Proceedings of the 14th Annual Conference on
  Computer Graphics and Interactive Techniques}}
  \emph{(\bibinfo{series}{SIGGRAPH `87})}.
\newblock


\bibitem[\protect\citeauthoryear{{Electronic Arts}}{{Electronic Arts}}{2019a}]%
        {battlefield-v}
\bibfield{author}{\bibinfo{person}{{Electronic Arts}}.}
  \bibinfo{year}{2019}\natexlab{a}.
\newblock \bibinfo{title}{Battlefield {V}}.
\newblock
\newblock


\bibitem[\protect\citeauthoryear{{Electronic Arts}}{{Electronic Arts}}{2019b}]%
        {picapica}
\bibfield{author}{\bibinfo{person}{{Electronic Arts}}.}
  \bibinfo{year}{2019}\natexlab{b}.
\newblock \bibinfo{title}{Project {PICA} {PICA}}.
\newblock
\newblock


\bibitem[\protect\citeauthoryear{Fascione, Hanika, Leone, Droske, Schwarzhaupt,
  Davidovi\v{c}, Weidlich, and Meng}{Fascione et~al\mbox{.}}{2018}]%
        {manuka}
\bibfield{author}{\bibinfo{person}{Luca Fascione}, \bibinfo{person}{Johannes
  Hanika}, \bibinfo{person}{Mark Leone}, \bibinfo{person}{Marc Droske},
  \bibinfo{person}{Jorge Schwarzhaupt}, \bibinfo{person}{Tom\'{a}\v{s}
  Davidovi\v{c}}, \bibinfo{person}{Andrea Weidlich}, {and}
  \bibinfo{person}{Johannes Meng}.} \bibinfo{year}{2018}\natexlab{}.
\newblock \showarticletitle{Manuka: A Batch-Shading Architecture for Spectral
  Path Tracing in Movie Production}.
\newblock \bibinfo{journal}{\emph{ACM Transactions on Graphics}}
  \bibinfo{volume}{37}, \bibinfo{number}{3}, Article \bibinfo{articleno}{31}
  (\bibinfo{date}{Aug.} \bibinfo{year}{2018}), \bibinfo{numpages}{18}~pages.
\newblock
\showISSN{0730-0301}


\bibitem[\protect\citeauthoryear{Georgiev, Ize, Farnsworth, Montoya-Vozmediano,
  King, Lommel, Jimenez, Anson, Ogaki, Johnston, Herubel, Russell, Servant, and
  Fajardo}{Georgiev et~al\mbox{.}}{2018}]%
        {arnold}
\bibfield{author}{\bibinfo{person}{Iliyan Georgiev}, \bibinfo{person}{Thiago
  Ize}, \bibinfo{person}{Mike Farnsworth}, \bibinfo{person}{Ram\'{o}n
  Montoya-Vozmediano}, \bibinfo{person}{Alan King}, \bibinfo{person}{Brecht~Van
  Lommel}, \bibinfo{person}{Angel Jimenez}, \bibinfo{person}{Oscar Anson},
  \bibinfo{person}{Shinji Ogaki}, \bibinfo{person}{Eric Johnston},
  \bibinfo{person}{Adrien Herubel}, \bibinfo{person}{Declan Russell},
  \bibinfo{person}{Fr{\'e}d{\'e}ric Servant}, {and} \bibinfo{person}{Marcos
  Fajardo}.} \bibinfo{year}{2018}\natexlab{}.
\newblock \showarticletitle{Arnold: A Brute-Force Production Path Tracer}.
\newblock \bibinfo{journal}{\emph{ACM Transactions on Graphics}}
  \bibinfo{volume}{37}, \bibinfo{number}{3}, Article \bibinfo{articleno}{32}
  (\bibinfo{date}{Aug.} \bibinfo{year}{2018}), \bibinfo{numpages}{12}~pages.
\newblock
\showISSN{0730-0301}


\bibitem[\protect\citeauthoryear{Hill, McAuley, Conty, Drobot, Heitz, Hery,
  Kulla, Lanz, Ling, Walster, Xie, Micciulla, and Villemin}{Hill
  et~al\mbox{.}}{2017}]%
        {sony-pbs}
\bibfield{author}{\bibinfo{person}{Stephen Hill}, \bibinfo{person}{Stephen
  McAuley}, \bibinfo{person}{Alejandro Conty}, \bibinfo{person}{Michal Drobot},
  \bibinfo{person}{Eric Heitz}, \bibinfo{person}{Christophe Hery},
  \bibinfo{person}{Christopher Kulla}, \bibinfo{person}{Jon Lanz},
  \bibinfo{person}{Junyi Ling}, \bibinfo{person}{Nathan Walster},
  \bibinfo{person}{Feng Xie}, \bibinfo{person}{Adam Micciulla}, {and}
  \bibinfo{person}{Ryusuke Villemin}.} \bibinfo{year}{2017}\natexlab{}.
\newblock \showarticletitle{Physically Based Shading in Theory and Practice}.
  In \bibinfo{booktitle}{\emph{ACM SIGGRAPH 2017 Courses}}
  \emph{(\bibinfo{series}{SIGGRAPH `17})}.
\newblock


\bibitem[\protect\citeauthoryear{Intel}{Intel}{2019}]%
        {oidn}
\bibfield{author}{\bibinfo{person}{Intel}.} \bibinfo{year}{2019}\natexlab{}.
\newblock \bibinfo{title}{Intel {Open} {Image} {Denoise}}.
\newblock \bibinfo{howpublished}{\url{https://openimagedenoise.github.io}}.
\newblock


\bibitem[\protect\citeauthoryear{Keim, Simmons, Teece, Reisweber, and
  Drakeley}{Keim et~al\mbox{.}}{2016}]%
        {disney-bonsai}
\bibfield{author}{\bibinfo{person}{Hans Keim}, \bibinfo{person}{Maryann
  Simmons}, \bibinfo{person}{Daniel Teece}, \bibinfo{person}{Jared Reisweber},
  {and} \bibinfo{person}{Sara Drakeley}.} \bibinfo{year}{2016}\natexlab{}.
\newblock \showarticletitle{Art-directable Procedural Vegetation in Disney's
  Zootopia}. In \bibinfo{booktitle}{\emph{ACM SIGGRAPH 2016 Talks}}
  \emph{(\bibinfo{series}{SIGGRAPH `16})}.
\newblock
\showISBNx{978-1-4503-4282-7}


\bibitem[\protect\citeauthoryear{Keller, Fascione, Fajardo, Georgiev,
  Christensen, Hanika, Eisenacher, and Nichols}{Keller et~al\mbox{.}}{2015}]%
        {keller_pt_course_2015}
\bibfield{author}{\bibinfo{person}{Alex Keller}, \bibinfo{person}{Luca
  Fascione}, \bibinfo{person}{Marcos Fajardo}, \bibinfo{person}{Iliyan
  Georgiev}, \bibinfo{person}{Per Christensen}, \bibinfo{person}{Johannes
  Hanika}, \bibinfo{person}{Christian Eisenacher}, {and}
  \bibinfo{person}{Gregory Nichols}.} \bibinfo{year}{2015}\natexlab{}.
\newblock \showarticletitle{The {Path} {Tracing} {Revolution} in the {Movie}
  {Industry}}. In \bibinfo{booktitle}{\emph{ACM SIGGRAPH 2015 Courses}}
  \emph{(\bibinfo{series}{SIGGRAPH `15})}.
\newblock


\bibitem[\protect\citeauthoryear{Kulla, Conty, Stein, and Gritz}{Kulla
  et~al\mbox{.}}{2018}]%
        {arnold-sony}
\bibfield{author}{\bibinfo{person}{Christopher Kulla},
  \bibinfo{person}{Alejandro Conty}, \bibinfo{person}{Clifford Stein}, {and}
  \bibinfo{person}{Larry Gritz}.} \bibinfo{year}{2018}\natexlab{}.
\newblock \showarticletitle{Sony Pictures Imageworks Arnold}.
\newblock \bibinfo{journal}{\emph{ACM Transactions on Graphics}}
  \bibinfo{volume}{37}, \bibinfo{number}{3}, Article \bibinfo{articleno}{29}
  (\bibinfo{date}{Aug.} \bibinfo{year}{2018}), \bibinfo{numpages}{18}~pages.
\newblock
\showISSN{0730-0301}


\bibitem[\protect\citeauthoryear{Lee, Green, Xie, and Tabellion}{Lee
  et~al\mbox{.}}{2017}]%
        {lee_vectorized_2017}
\bibfield{author}{\bibinfo{person}{Mark Lee}, \bibinfo{person}{Brian Green},
  \bibinfo{person}{Feng Xie}, {and} \bibinfo{person}{Eric Tabellion}.}
  \bibinfo{year}{2017}\natexlab{}.
\newblock \showarticletitle{Vectorized {{Production Path Tracing}}}. In
  \bibinfo{booktitle}{\emph{Proceedings of High Performance Graphics}}
  \emph{(\bibinfo{series}{HPG `17})}.
\newblock


\bibitem[\protect\citeauthoryear{Mara, McGuire, Bitterli, and Jarosz}{Mara
  et~al\mbox{.}}{2017}]%
        {mara_denoise_2017}
\bibfield{author}{\bibinfo{person}{Michael Mara}, \bibinfo{person}{Morgan
  McGuire}, \bibinfo{person}{Benedikt Bitterli}, {and}
  \bibinfo{person}{Wojciech Jarosz}.} \bibinfo{year}{2017}\natexlab{}.
\newblock \showarticletitle{An Efficient Denoising Algorithm for Global
  Illumination}. In \bibinfo{booktitle}{\emph{Proceedings of High Performance
  Graphics}} \emph{(\bibinfo{series}{HPG `17})}.
\newblock


\bibitem[\protect\citeauthoryear{Parker, Bigler, Dietrich, Friedrich, Hoberock,
  Luebke, McAllister, McGuire, Morley, Robison, and Stich}{Parker
  et~al\mbox{.}}{2010}]%
        {optix}
\bibfield{author}{\bibinfo{person}{Steven~G. Parker}, \bibinfo{person}{James
  Bigler}, \bibinfo{person}{Andreas Dietrich}, \bibinfo{person}{Heiko
  Friedrich}, \bibinfo{person}{Jared Hoberock}, \bibinfo{person}{David Luebke},
  \bibinfo{person}{David McAllister}, \bibinfo{person}{Morgan McGuire},
  \bibinfo{person}{Keith Morley}, \bibinfo{person}{Austin Robison}, {and}
  \bibinfo{person}{Martin Stich}.} \bibinfo{year}{2010}\natexlab{}.
\newblock \showarticletitle{{OptiX}: {A} {General} {Purpose} {Ray} {Tracing}
  {Engine}}.
\newblock \bibinfo{journal}{\emph{ACM Transactions on Graphics (Proceedings of
  ACM SIGGRAPH)}}.
\newblock


\bibitem[\protect\citeauthoryear{Pharr}{Pharr}{2018a}]%
        {pharr_tog_intro}
\bibfield{author}{\bibinfo{person}{Matt Pharr}.}
  \bibinfo{year}{2018}\natexlab{a}.
\newblock \showarticletitle{Guest {Editor's} {Introduction}: {Special} {Issue}
  on {Production} {Rendering}}.
\newblock \bibinfo{journal}{\emph{ACM Transactions on Graphics}}
  \bibinfo{volume}{37}, \bibinfo{number}{3}, Article \bibinfo{articleno}{28}
  (\bibinfo{date}{July} \bibinfo{year}{2018}), \bibinfo{numpages}{4}~pages.
\newblock
\showISSN{0730-0301}


\bibitem[\protect\citeauthoryear{Pharr}{Pharr}{2018b}]%
        {matt_pharr_blog}
\bibfield{author}{\bibinfo{person}{Matt Pharr}.}
  \bibinfo{year}{2018}\natexlab{b}.
\newblock \bibinfo{title}{Swallowing the {Elephant}}.
\newblock
  \bibinfo{howpublished}{\url{https://pharr.org/matt/blog/2018/07/16/moana-island-pbrt-all.html}}.
\newblock


\bibitem[\protect\citeauthoryear{Pharr, Jakob, and Humphreys}{Pharr
  et~al\mbox{.}}{2016}]%
        {pbrt}
\bibfield{author}{\bibinfo{person}{Matt Pharr}, \bibinfo{person}{Wenzel Jakob},
  {and} \bibinfo{person}{Greg Humphreys}.} \bibinfo{year}{2016}\natexlab{}.
\newblock \showarticletitle{Physically Based Rendering: From Theory to
  Implementation (3rd ed.)}.
\newblock  (\bibinfo{year}{2016}), \bibinfo{pages}{1200}.
\newblock


\bibitem[\protect\citeauthoryear{Pharr and Mark}{Pharr and Mark}{2012}]%
        {ispc}
\bibfield{author}{\bibinfo{person}{Matt Pharr} {and} \bibinfo{person}{Bill
  Mark}.} \bibinfo{year}{2012}\natexlab{}.
\newblock \showarticletitle{{ISPC}: {A} {SPMD} {Compiler} for
  {High}-{Performance} {CPU} {Programming}}. In
  \bibinfo{booktitle}{\emph{Proceedings of Innovative Parallel Computing
  (inPar)}}.
\newblock


\bibitem[\protect\citeauthoryear{Poh, Kilgore, Wichitscripornkul, and
  Monheit}{Poh et~al\mbox{.}}{2018}]%
        {poh_2018}
\bibfield{author}{\bibinfo{person}{Kiki Poh}, \bibinfo{person}{Michael
  Kilgore}, \bibinfo{person}{Tom Wichitscripornkul}, {and}
  \bibinfo{person}{Gary Monheit}.} \bibinfo{year}{2018}\natexlab{}.
\newblock \showarticletitle{Using USD Shading to Provide the ``Extra'' Touch on
  Incredibles2}. In \bibinfo{booktitle}{\emph{ACM SIGGRAPH 2018 Talks}}
  \emph{(\bibinfo{series}{SIGGRAPH `18})}.
\newblock


\bibitem[\protect\citeauthoryear{Schied, Kaplanyan, Wyman, Patney, Chaitanya,
  Burgess, Liu, Dachsbacher, Lefohn, and Salvi}{Schied et~al\mbox{.}}{2017}]%
        {schied_denoise_2017}
\bibfield{author}{\bibinfo{person}{Christoph Schied}, \bibinfo{person}{Anton
  Kaplanyan}, \bibinfo{person}{Chris Wyman}, \bibinfo{person}{Anjul Patney},
  \bibinfo{person}{Chakravarty R.~Alla Chaitanya}, \bibinfo{person}{John
  Burgess}, \bibinfo{person}{Shiqiu Liu}, \bibinfo{person}{Carsten
  Dachsbacher}, \bibinfo{person}{Aaron Lefohn}, {and} \bibinfo{person}{Marco
  Salvi}.} \bibinfo{year}{2017}\natexlab{}.
\newblock \showarticletitle{Spatiotemporal Variance-guided Filtering: Real-time
  Reconstruction for Path-traced Global Illumination}. In
  \bibinfo{booktitle}{\emph{Proceedings of High Performance Graphics}}
  \emph{(\bibinfo{series}{HPG `17})}.
\newblock


\bibitem[\protect\citeauthoryear{Tamstorf and Pritchett}{Tamstorf and
  Pritchett}{2018}]%
        {moana-whitepaper}
\bibfield{author}{\bibinfo{person}{Rasmus Tamstorf} {and}
  \bibinfo{person}{Heather Pritchett}.} \bibinfo{year}{2018}\natexlab{}.
\newblock \bibinfo{title}{Moana Island Scene}.
\newblock
  \bibinfo{howpublished}{\url{http://datasets.disneyanimation.com/moanaislandscene/island-README-v1.1.pdf}}.
\newblock


\bibitem[\protect\citeauthoryear{Usher, Wald, Amstutz, Günther, Brownlee, and
  Pascucci}{Usher et~al\mbox{.}}{2019}]%
        {usher_scalable_2019}
\bibfield{author}{\bibinfo{person}{Will Usher}, \bibinfo{person}{Ingo Wald},
  \bibinfo{person}{Jefferson Amstutz}, \bibinfo{person}{Johannes Günther},
  \bibinfo{person}{Carson Brownlee}, {and} \bibinfo{person}{Valerio Pascucci}.}
  \bibinfo{year}{2019}\natexlab{}.
\newblock \showarticletitle{{Scalable Ray Tracing Using the Distributed
  FrameBuffer}}.
\newblock \bibinfo{journal}{\emph{Computer Graphics Forum}}
  (\bibinfo{year}{2019}).
\newblock


\bibitem[\protect\citeauthoryear{Wald, Benthin, and Slusallek}{Wald
  et~al\mbox{.}}{2002}]%
        {wald_flexible_2002}
\bibfield{author}{\bibinfo{person}{Ingo Wald}, \bibinfo{person}{Carsten
  Benthin}, {and} \bibinfo{person}{Philipp Slusallek}.}
  \bibinfo{year}{2002}\natexlab{}.
\newblock \bibinfo{booktitle}{\emph{A {{Flexible}} and {{Scalable Rendering
  Engine}} for {{Interactive 3D Graphics}}}}.
\newblock \bibinfo{type}{{T}echnical {R}eport}. \bibinfo{institution}{{Saarland
  University}}.
\newblock


\bibitem[\protect\citeauthoryear{Wald, Johnson, Amstutz, Brownlee, Knoll,
  Jeffers, G\"unther, and Navr\'atil}{Wald et~al\mbox{.}}{2017}]%
        {ospray}
\bibfield{author}{\bibinfo{person}{Ingo Wald}, \bibinfo{person}{Greg~P.
  Johnson}, \bibinfo{person}{Jefferson Amstutz}, \bibinfo{person}{Carson
  Brownlee}, \bibinfo{person}{Aaron Knoll}, \bibinfo{person}{Jim Jeffers},
  \bibinfo{person}{Johannes G\"unther}, {and} \bibinfo{person}{Paul
  Navr\'atil}.} \bibinfo{year}{2017}\natexlab{}.
\newblock \showarticletitle{{{OSPRay}} - {{A CPU Ray Tracing Framework}} for
  {{Scientific Visualization}}}.
\newblock \bibinfo{journal}{\emph{IEEE Transactions on Visualization and
  Computer Graphics}} (\bibinfo{year}{2017}).
\newblock


\bibitem[\protect\citeauthoryear{Wald, Woop, Benthin, Johnson, and Ernst}{Wald
  et~al\mbox{.}}{2014}]%
        {embree}
\bibfield{author}{\bibinfo{person}{Ingo Wald}, \bibinfo{person}{Sven Woop},
  \bibinfo{person}{Carsten Benthin}, \bibinfo{person}{Gregory~S. Johnson},
  {and} \bibinfo{person}{Manfred Ernst}.} \bibinfo{year}{2014}\natexlab{}.
\newblock \showarticletitle{Embree: A Kernel Framework for Efficient {CPU} Ray
  Tracing}.
\newblock \bibinfo{journal}{\emph{ACM Transactions on Graphics (Proceedings of
  ACM SIGGRAPH)}}  \bibinfo{volume}{33} (\bibinfo{year}{2014}).
\newblock


\end{thebibliography}
\end{document}